\documentclass{pnastwo}

\pdfoutput=1
\usepackage{pnastwof}
\usepackage{color,graphicx}
\usepackage[resetlabels]{multibib}
\usepackage{array} 
\usepackage{url}
\usepackage{xcolor}
\usepackage{tabularx}
\usepackage{amssymb,amsfonts,amsmath}
\usepackage[comma,sort&compress]{natbib}
\def \aap{A\&A}

\def \apjl{ApJ}
\def \apjs{ApJS}

\makeatletter
\def\tabular{\global\setbox\tablewide\hbox\bgroup
\let\@halignto\@empty\@tabular}

\def\endtabular{\crcr\egroup\egroup $\egroup\egroup
\centerline{\vbox{\hsize\wd\tablewide 
\currtabcaption\vskip1pt
}}
\dimen0=\wd\tablewide
\centerline{\hbox{\unhbox\tablewide}}
\centerline{\vtop{\hsize=\dimen0 \tablenotes}}
\global\let\currtabcaption\relax}

\makeatother

\definecolor{lightblue}{rgb}{0.80,0.85,0.92}
\definecolor{darkblue}{rgb}{0.00,0.41,0.66}

\begin{document}

\title{Prospects for Detecting Oxygen, Water, and Chlorophyll on an Exo-Earth}

\author{Timothy D.~Brandt\affil{1}{School of Natural Sciences, Institute for Advanced Study, Princeton, NJ, USA.} \and David S.~Spiegel\affil{1}{School of Natural Sciences, Institute for Advanced Study, Princeton, NJ, USA.}}
\footlineauthor{Brandt and Spiegel}

\contributor{Submitted to the Proceedings of the National Academy of Sciences of the United States of America on \today{}.}
\maketitle

\begin{article}

\begin{abstract}
The goal of finding and characterizing nearby Earth-like planets is driving many NASA high-contrast flagship mission concepts, the latest of which is known as the Advanced Technology Large-Aperture Space Telescope (ATLAST).  In this article, we calculate the optimal spectral resolution $R=\lambda/\delta\lambda$ and minimum signal-to-noise ratio per spectral bin (SNR), two central design requirements for a high-contrast space mission, in order to detect signatures of water, oxygen, and chlorophyll on an Earth twin.  We first develop a minimally parametric model and demonstrate its ability to fit synthetic and observed Earth spectra; this allows us to measure the statistical evidence for each component's presence.  We find that water is the easiest to detect, requiring a resolution $R \gtrsim 20$, while the optimal resolution for oxygen is likely to be closer to $R = 150$, somewhat higher than the canonical value in the literature.  At these resolutions, detecting oxygen will require $\sim$2 times the SNR as water.  Chlorophyll requires $\sim$6 times the SNR as oxygen for an Earth twin, only falling to oxygen-like levels of detectability for a low cloud cover and/or a large vegetation covering fraction.  This suggests designing a mission for sensitivity to oxygen and adopting a multi-tiered observing strategy, first targeting water, then oxygen on the more favorable planets, and finally chlorophyll on only the most promising worlds.  

\end{abstract}


\noindent\fcolorbox{darkblue}{lightblue}{\parbox{\dimexpr \linewidth-2\fboxsep-2\fboxrule}{
\abstractfont \color{darkblue} { Significance\\[5pt]} 
One of NASA's most important long-term goals is to detect and characterize terrestrial exoplanets, and to search their spectra for signs of life.  This overarching goal is currently driving concepts for a future high-contrast flagship mission.  We determine the fidelity with which such a mission would need to measure an exo-Earth's spectrum in order to detect oxygen, water, and chlorophyll.  Our results suggest that a well-designed space mission could detect O$_2$ and H$_2$O in a nearby Earth twin, but that it would need to be significantly more sensitive (or very lucky) to see chlorophyll.  We suggest designing the instrument with an eye towards oxygen, and perhaps looking for chlorophyll around one or a few exceptional targets.}}

\section{Introduction}

While indirect methods have now discovered several thousand exoplanets \cite{Schneider+Dedieu+Sidaner+etal2011, Wright+Fakhouri+Marcy+etal2011, Rein2012, Akeson+Chen+Ciardi+etal_2013}, most of these are currently inaccessible to characterization.  Direct imaging offers the ability to observe an exoplanet in either thermal or reflected light, and provides a window into the structure and composition of its atmosphere.  Ultimately, one of NASA's goals is to find and characterize terrestrial exoplanets around nearby stars and to search for molecules and biosignatures.  The {\it Terrestrial Planet Finder} \citep[TPF, e.g.][]{Beichman2003, Levine+Shaklan+Kasting+etal_2009} was one mission concept with this goal in mind, while the {\it Advanced Technology Large-Aperture Space Telescope} \citep[ATLAST, ][]{Postman+Argabright+Arnold+etal_2009} represents a more recent proposal.
Such a mission would target at least the optical wavelength range from $\sim$0.5 to $\sim$1 $\mu$m in order to look for spectral indications of molecular oxygen (O$_2$), ozone (O$_3$), and water (H$_2$O).  These molecules are the most prominent absorbers in this spectral range, and are all critical species for terrestrial life.  

Several studies have looked at the prospects for detecting biosignatures on an Earth twin \cite{Desmarais+Harwit+Jucks+etal2002, Kaltenegger+Selsis2007, Davies+Benner+Cleland+etal2009}, or for learning about an exo-Earth's surface from phase variations in its colors \citep{Fujii+Kawahara+Suto+etal_2010, Fujii+Kawahara+Suto+etal_2011, Cowan+Robinson+Livengood+etal2011, Cowan+Abbot+Voigt2012}.  Of course, there is no unique exo-Earth spectrum: the composition of Earth's atmosphere, including its oxygen abundance, has changed enormously throughout life's existence \citep{Kaltenegger+Traub+Jucks_2007}.  Furthermore, features that are biosignatures in Earth's spectrum may not necessarily be so in a terrestrial exoplanet.  For instance, diatomic oxygen can, under some circumstances, also be produced abiotically \citep{Wordsworth+Pierrehumbert_2014}.  
More speculatively, chlorophyll shows a strong increase in its albedo around 0.7 $\mu$m (the ``red edge''), which could be detected on an exo-Earth \cite{Seager+Turner+Schafer+etal2005, Montanes-Rodriguez+Palle+Goode+etal_2006}.  Such an argument relies on the uniqueness of the chlorophyll family of molecules as the basis for photosynthesis \cite{Kiang+Siefert+Govindjee+etal2007}.  

Many of the papers referenced above have run detailed model atmosphere calculations.  Our goal here is different.  We seek to construct the {\it simplest} model that can adequately reproduce a terrestrial planet's spectrum, and to use it to derive statistically rigorous criteria to claim detections of molecular species.  
Using a model with few free parameters increases the statistical significance with which the most interesting parameters (like the H$_2$O or O$_2$ column) may be estimated.
We make as few assumptions as possible about the (highly uncertain) performance of a future high-contrast space mission.  Rather than working from an instrument to detectability, we turn the problem around, and attempt to quantify the optimal design and minimum performance needed to reach NASA's terrestrial planet characterization goals.

\section{Terrestrial Planet Spectra}

We begin with a crude, but roughly correct, approximation of a terrestrial planet spectrum in reflection.  Our goal is to capture the main spectral features of Earth in a context where we can easily modify the cloud, surface, and atmospheric compositions, allowing us to validate the statistical approach we present in this paper.

We assume a (wavelength-dependent) surface albedo $\alpha_\lambda$, a cloud albedo $c_\lambda$, a cloud fraction $f_c$, an optical depth to Rayleigh scattering $\mathcal{R}_\lambda$ (assumed to be small), and absorption cross-sections $\sigma_s[\lambda]$ for chemical species $s$ in Earth's atmosphere.  The wavelength-dependent optical depth for a species $s$ is the product of its cross-section $\sigma_s[\lambda]$ and its column density $N_s$, so that $\tau_\lambda$, the full atmosphere's optical depth, is
\begin{equation}
\tau_\lambda = \sum_{s} \sigma_s[\lambda] N_s~, 
\end{equation}
where the sum over $s$ includes the molecules O$_2$, O$_3$, and H$_2$O.  

\begin{figure}
\noindent
\centering
\includegraphics[width=\linewidth]{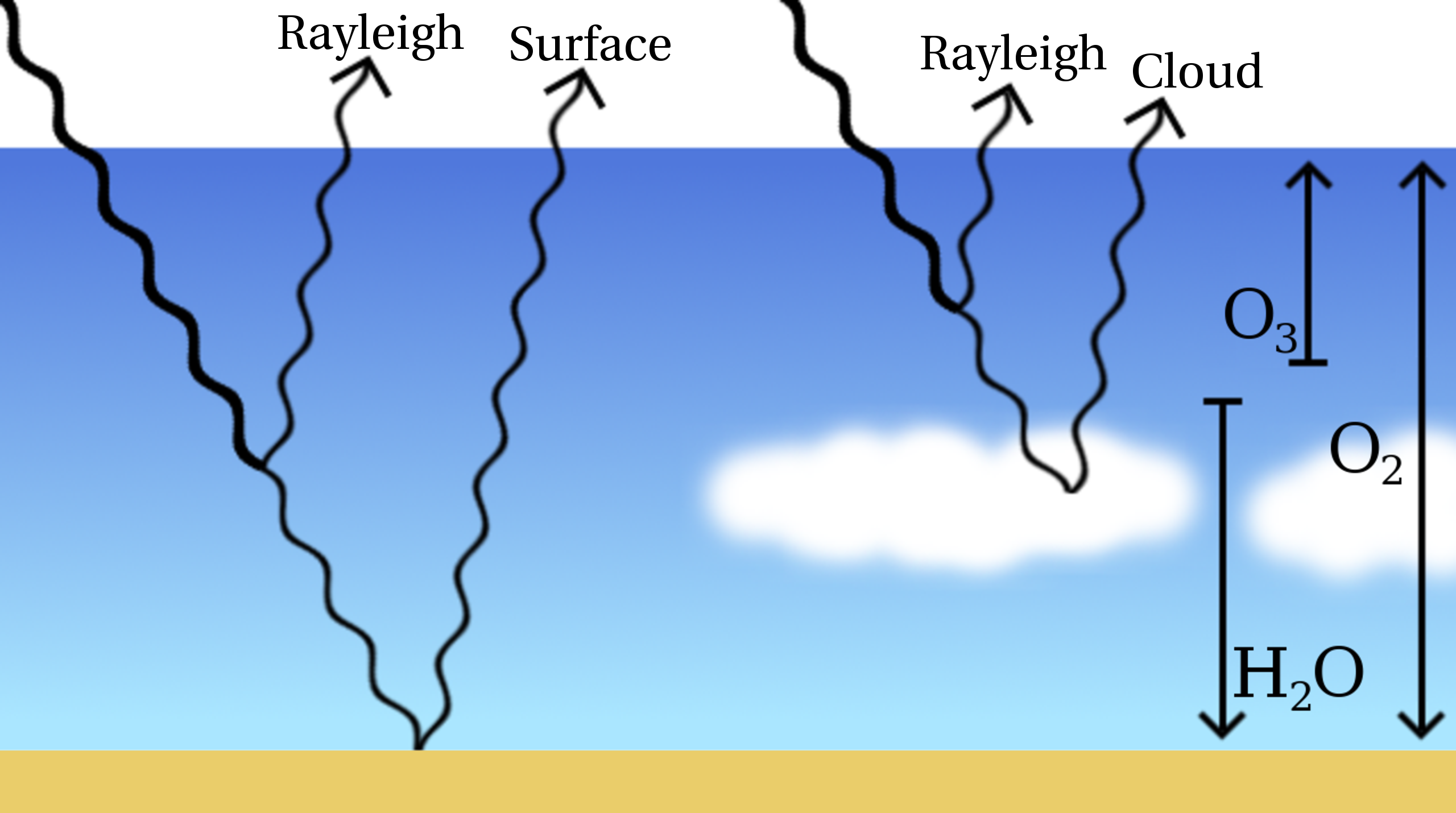}
\caption{Illustration of the scattering approximation described by Equation \eqref{eq:full_atm}; we approximate all Rayleigh scattering as occurring below half of the available atmosphere.  We assume that all photons pass through the stratospheric O$_3$ layer twice, while photons scattered by clouds and Rayeigh scattered by atmospheric O$_2$ and N$_2$ pass through less than half the water vapor column.
\label{fig:twostreammodel}}
\end{figure}

We approximate all Rayleigh scattering as occurring through half the available atmosphere, as illustrated in Figure \ref{fig:twostreammodel}.  While O$_2$ is relatively well-mixed up to the stratosphere, most O$_3$ is above, and most water below, the tropopause \cite{Brewer_1949}.  We therefore assume that {\it all} scattered photons pass through the entire O$_3$ column twice, while photons Rayleigh scattered by O$_2$ and N$_2$, as well as those scattered by clouds, pass through less than half the water vapor column.  We find that assuming these photons to pass through 20\% of the H$_2$O column reproduces the approximate strength of the water features in more detailed models \cite{Robinson+Meadows+Crisp+etal_2011}.

Our full approximation to the reflected flux density becomes
\begin{align}
\frac{F_{\rm refl}}{F_{\star}} \approx &\frac{\cal R_\lambda}{2} f_c 
\exp\left[-2\tau_{\rm top} \right]
+ \left( 1 - \frac{\cal R_\lambda}{2} \right) f_c c_\lambda \exp\left[-2\tau_{\rm mid}\right] \nonumber \\
&+ {\cal R_\lambda} \left( 1 - f_c \right) \exp\left[-2\tau_{\rm mid} \right] \nonumber \\
&+ \left( 1 - {\cal R_\lambda} \right) \left( 1 - f_c \right) \alpha_\lambda \exp\left[-2\tau_{\rm full} \right]~,
\label{eq:full_atm}
\end{align}
where $F_{\star}$ is the incident (Solar\footnote{http://rredc.nrel.gov/solar/spectra/am0/}) flux on the planetary atmosphere, and $F_{\rm refl}$ is the reflected flux.  The first term approximates Rayleigh scattering above the clouds, the second term accounts for scattering by the clouds themselves, the third term is Rayleigh scattering above the surface, and the last term is scattering by the surface (we have dropped all terms with ${\cal R}_\lambda^2$).  The three optical depths, through $1/4$, $1/2$, and all of the atmosphere, are given by
\begin{align}
\tau_{\rm top} &= \sigma_{\rm O3} N_{\rm O3} + \sigma_{\rm O2} N_{\rm O2}/4 + \sigma_{\rm H2O} N_{\rm H2O}/10~, \\
\tau_{\rm mid} &= \sigma_{\rm O3} N_{\rm O3} + \sigma_{\rm O2} N_{\rm O2}/2 + \sigma_{\rm H2O} N_{\rm H2O}/5~,\quad {\rm and} \\
\tau_{\rm full} &= \sigma_{\rm O3} N_{\rm O3} + \sigma_{\rm O2} N_{\rm O2} + \sigma_{\rm H2O} N_{\rm H2O}~. 
\end{align}

We use the ASTER spectral library \citep{Baldridge+Hook+Grove+etal_2009} for our surface albedos excepting water, an approximation based on \cite{McLinden+McConnell+Griffioen+etal_1997} for oceans (including a substantial correction for specular reflection at blue wavelengths), \cite{Kokhanovsky_2004} for water cloud albedos, and \cite{Rugheimer+Kaltenegger+Zsom+etal_2013} for the cross-sections of O$_2$, H$_2$O, and O$_3$.  We assume the surface to be 70\% water, 10\% sand, 10\% vegetation, 5\% dry grass, and 5\% snow, consistent with estimates from data taken by the {\it MODIS} satellite \citep{Salomonson+Barnes+Maymon+etal_1989} and tabulated in \cite{Fujii+Kawahara+Suto+etal_2010}. 
We take our normalization of the Rayleigh scattering optical depths for Earth's atmosphere from \cite{Froehlich+Shaw_1980}.  We take a 50\% cloud coverage, noting that its main effects are to obscure spectral features from the surface and dilute the water features (cloud albedos are gray to an excellent approximation, while photons must still travel through at least $\sim$1/2 the atmosphere).  The effective albedo of clouds is $\sim$60--65\% across the wavelength range, washing out surface spectral features by a factor $\gg$$1$ when the clouds are optically thick.

We create mock spectra by smoothing the output of Equation \eqref{eq:full_atm} to an adopted spectral resolution using a Gaussian line-spread function,
\begin{equation}
{\rm LSF}[\lambda/\lambda_0] \propto \exp\left[ -(4 \ln 2) R^2 \left( \frac{\lambda}{\lambda_0} - 1 \right)^2 \right]~,
\label{eq:lsf}
\end{equation}
where $R$ is the (dimensionless) instrumental resolution, the full width at half maximum of the Gaussian.  We assume that these spectra are well-sampled (i.e.~at the Nyquist rate or better, $\delta \lambda \lesssim \lambda/(2R)$).
Binning the spectra, in contrast, would be equivalent to convolving with a boxcar line-spread function and sampling at the boxcar's full width.  Such undersampling makes spectral reconstruction dependent on excellent knowledge of the line-spread function and even on the centers chosen (often arbitrarily) for the wavelength bins.  

Finally, we convert our units into flux density $f_\nu$, proportional to photons per logarithmic wavelength bin.  The resulting flux density is approximately constant (within a factor of $\sim$1.5) across our wavelength range: an upturn in albedo from Rayleigh scattering at $\lesssim$0.5 $\mu$m compensates for a sharp fall in stellar intensity.  The Supporting Information provides additional details.  We further assume that noise, whether from speckle residuals, background photon noise, read noise, or some other source, is also a constant across the wavelength range, so that the signal-to-noise ratio per $\lambda/R$ bin (SNR) is nearly independent of wavelength (apart from the centers of spectral lines/features, where the signal drops).  

\begin{figure}[t]
\vspace{-0.035\linewidth}
\noindent
\includegraphics[width=\linewidth]{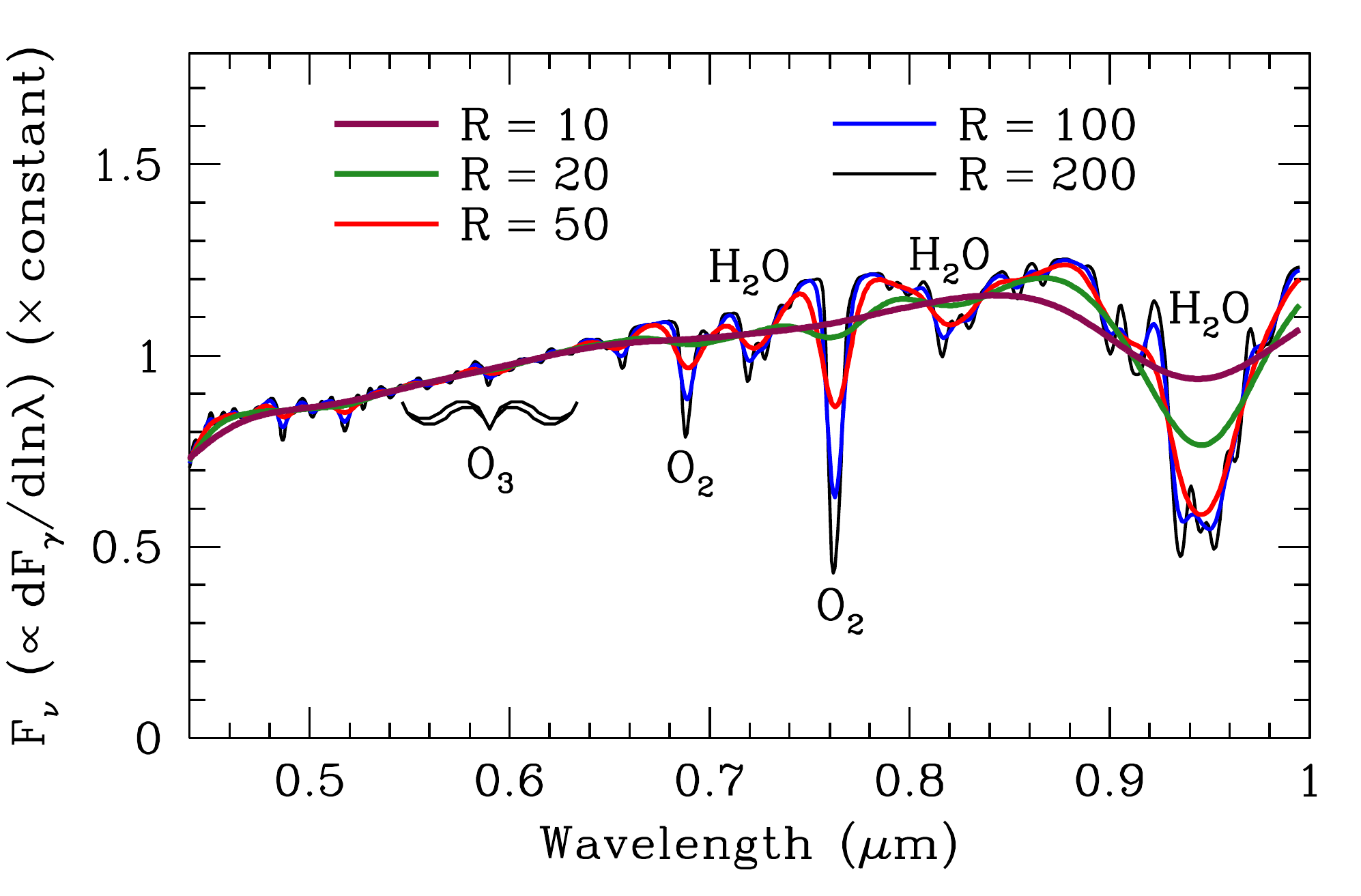}
\caption{Comparison of our spectrum of an Earth twin convolved to a given spectral resolution with a Gaussian line-spread function.  The prominent O$_2$ absorption feature at 0.76 $\mu$m becomes completely blended with neighboring water features for $R \lesssim 20$, while the O$_3$ feature is wide and shallow, and very difficult to see.
\label{fig:mock_spectra}
}
\end{figure}

Figure \ref{fig:mock_spectra}
shows noiseless mock spectra of an Earth twin at full phase for a variety of spectral resolutions.  For spectral resolutions $R \lesssim 20$, the prominent 0.76 $\mu$m O$_2$ absorption feature becomes almost completely blended with neighboring water features, making the spectrum flat.  The ozone band centered at $\lambda \sim 0.59$ $\mu$m is both broad and shallow, making it difficult to see at any spectral resolution.  Water features, on the other hand, remain conspicuous down to $R \sim 20$.

\section{A Minimally Parametric Model}

We assume 
that the spectrum of a hypothetical terrestrial exoplanet will be modeled using a 
known stellar spectrum, absorption spectra of possible atmospheric constituents, and Rayleigh scattering.  Unfortunately, a space mission capable of achieving contrasts of $10^{-10}$ will be unlikely to have perfect spectrophotometric calibration of a faint exoplanet.  Surface materials have spectral albedos that vary across the visible wavelength range, though most plausible surface materials, including soils, snow, and water, lack sharp spectral features from $\sim$0.5 to 1 $\mu$m.  
We therefore combine the unknown spectrophotometric calibration and surface albedos into a free multiplicative low-order polynomial.  
We make one exception: the ``red edge'' of chlorophyll at $\sim$ 0.72 $\mu$m, which we approximate using a softened Heaviside function,
\begin{equation}
H[\lambda] = \left(1 + \exp\left[ 80(0.72 - \lambda/\mu {\rm m}) \right] \right)^{-1} ~,
\label{eq:heaviside}
\end{equation}
with the center and width chosen to match the feature in vegetation in the ASTER libraries.  

Our model for a terrestrial planet's reflection spectrum takes the form
\begin{align}
{F_\lambda} \approx {F_{\lambda, \star}}
\left( \sum_i a_i \lambda^i \right) \Bigg( &\left(1 + b H[\lambda] \right)
\exp \left[ -\sum_{s} 2 \sigma_s[\lambda] N_{s} \right] \nonumber \\
&+ \frac{c}{\lambda^4} \exp \left[ -\sum_{s} \sigma_s[\lambda] N_{s} \right]
\Bigg)~,
\label{eq:mpm}
\end{align}
where $\sigma_s[\lambda]$ is the (known) cross-section of a molecular species $s$, $N_{s}$ its column density (making $\sigma_s[\lambda] N_s$ the optical depth of each pass through the atmosphere), and the first term, a polynomial, accounts for uncertainties in the surface (and cloud) albedo and spectrophotometric calibration.
We crudely include Rayleigh scattering with a term $c/\lambda^4$, making the approximation that all Rayleigh scattering occurs beneath half the atmosphere.  Photons scattered off the surface pass through the atmosphere twice, accounting for the factor of 2 in the first sum over species $s$.  We caution against interpreting the Rayleigh-like term too literally; it also combines with the polynomial term to add flexibility to account for the unknown surface albedo.  The vegetation parameter $b$, which we can either fix to be zero or allow to float, approximates the addition of an arbitrary vegetation covering fraction.  We compute Equation \eqref{eq:mpm} at very high spectral resolution and then convolve it with the line-spread function (Equation \eqref{eq:lsf}).

We fit Equation \eqref{eq:mpm} to our mock terrestrial planet spectra, varying its parameters to minimize
\begin{equation}
\chi^2 = \sum_\lambda \frac{\left( F_{\lambda,\,{\rm obs}} - F_{\lambda,\,{\rm model}} \right)^2}{\sigma^2_\lambda}~,
\end{equation}
where $\sigma_\lambda$ is the measurement error at $\lambda$.  We fit as many as eight free parameters:
a column density $N_s \geq 0$ for each of O$_2$, O$_3$, and H$_2$O, a normalization $c$ of Rayleigh scattering, three polynomial coefficients $a_i$ (for a free quadratic), and a chlorophyll strength $b$.  By fixing one (or more) of the $N_s$ to be zero, the minimum value of $\chi^2$ will generally increase as the model loses flexibility; the magnitude of the increase is a measure of the evidence for that species' presence ($N_s > 0$).  

\begin{figure}[t]
\noindent
\includegraphics[width=\linewidth]{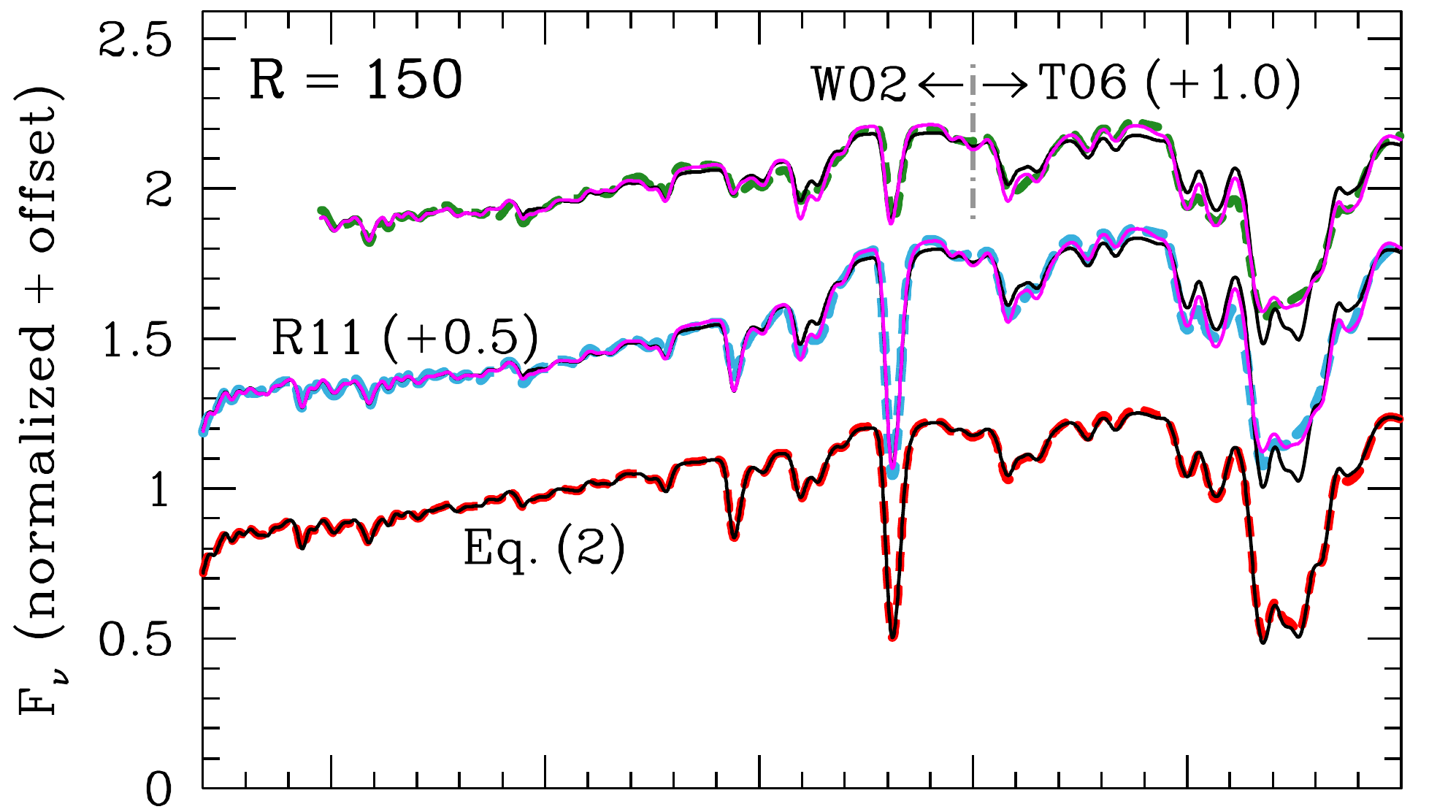}
\includegraphics[width=\linewidth]{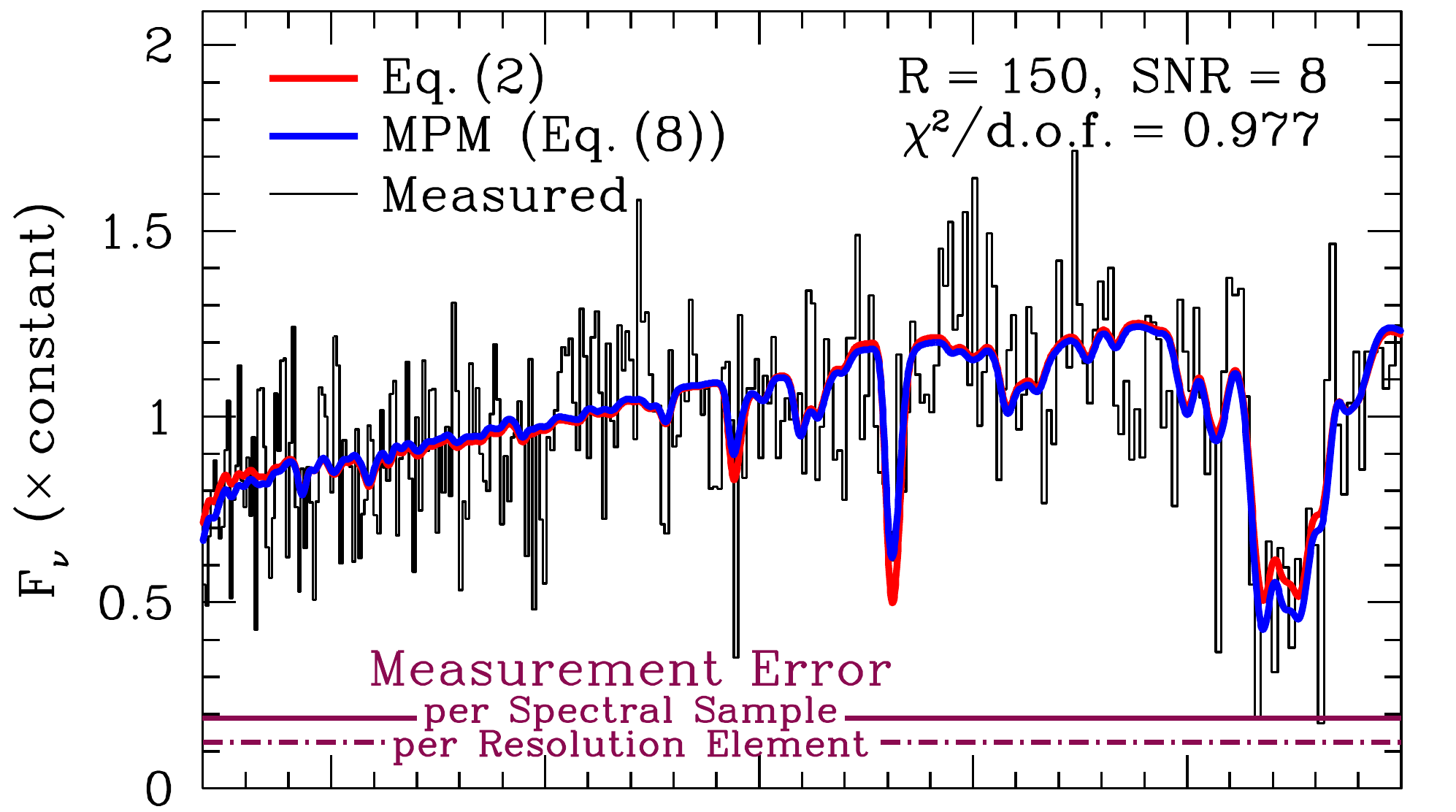}
\includegraphics[width=\linewidth]{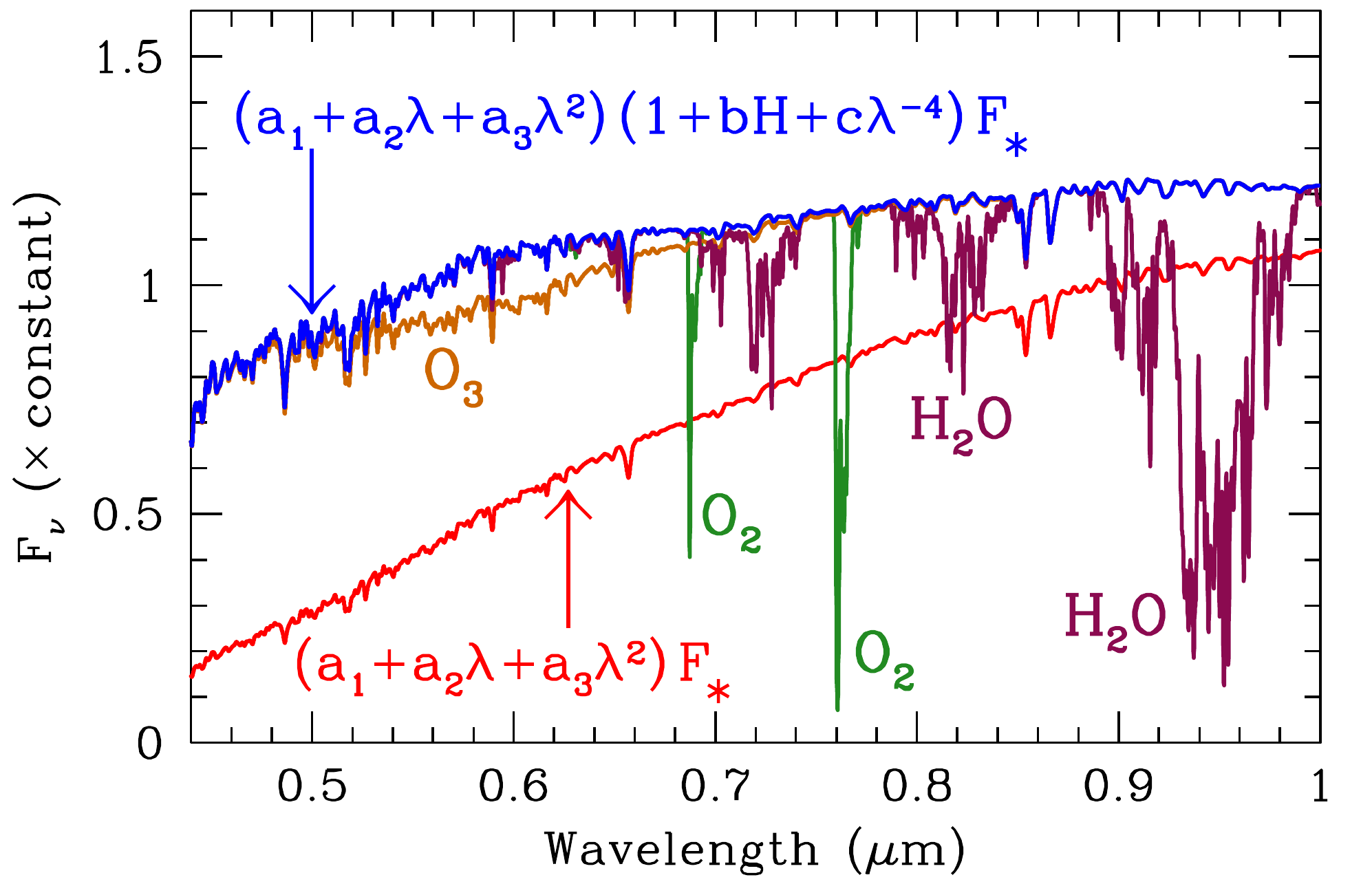}
\caption{Top panel: the best-fitting minimally parametric model (Equation \eqref{eq:mpm}, black curves) to Equation \eqref{eq:full_atm} (red dashed curve), a VPL model \cite{Robinson+Meadows+Crisp+etal_2011} (R11, blue dashed curve), and a spliced Earthshine spectrum \cite{Woolf+Smith+Traub+etal_2002, Turnbull+Traub+Jucks+etal_2006} (W02+T06, green dashed curve).  The magenta curves add one additional parameter to Equation \eqref{eq:mpm}, the fraction of the atmosphere containing water vapor, and provide a much better fit to the deep water features.  Middle panel: the fit of our minimally parametric model (MPM, Equation \eqref{eq:mpm}, blue curve) to one realization of Gaussian noise with ${\rm SNR}=8$ (black histogram).  Bottom panel: the fit in the middle panel decomposed by terms, at $R=1000$.  We caution against interpreting the `Rayleigh' term ($\propto \lambda^{-4}$) too literally, as this term ends up including much of the spectral variation in albedo.
\label{fig:demo}
}
\end{figure}

We test our minimally parametric model, Equation \eqref{eq:mpm}, against three reference spectra: the output of Equation \eqref{eq:full_atm}; a full radiative transfer model from the Virtual Planet Laboratory (VPL) \cite{Robinson+Meadows+Crisp+etal_2011}; and an Earthshine spectrum spliced together from visible and near-infrared observations \cite{Woolf+Smith+Traub+etal_2002, Turnbull+Traub+Jucks+etal_2006}.  The top panel of Figure \ref{fig:demo} shows these fits at a spectral resolution $R = 150$; the thick, dashed, colored lines are the reference spectra, while the thin black lines are the fits using Equation \eqref{eq:mpm}.  Equation \eqref{eq:mpm} has difficulty reproducing the exact shape of the deep water features.  If we extend the model by confining the water vapor to a fraction $f$ of the atmosphere (adding one additional free parameter, for a new total of 9), the fits improve significantly (thin magenta curves).

We say that Equation \eqref{eq:mpm} becomes inadequate when its best fit $\chi^2$ exceeds the number of degrees of freedom by the same amount required to detect an atmospheric feature ($\Delta \chi^2 \sim 10$, as we derive in the following section).  In other words, this is the point at which the information missing in Equation \eqref{eq:mpm} might be enough to detect an additional feature of the planet's spectrum.  According to this criterion, at $R = 150$, Equation \eqref{eq:mpm} provides a satisfactory fit to Equation \eqref{eq:full_atm} for ${\rm SNR} \lesssim 45$, and to the VPL model and the spliced Earthshine spectrum for ${\rm SNR} \lesssim 10$.  Adding one additional parameter, the fraction of the atmosphere free of water, enables the model to fit the VPL and the Earthshine spectra up to ${\rm SNR} \sim 20$ at $R=150$.  In all cases, our neglect of non-gray surface albedos and detailed radiative transfer appears to be unimportant except at very high SNR.

The middle panel of Figure \ref{fig:demo} shows the fit of Equation \eqref{eq:mpm} (blue curve) to Equation \eqref{eq:full_atm} (red curve) at $R=150$ for one realization of Gaussian noise with ${\rm SNR}=8$ (black histogram).  The noisy spectrum is sampled 
at $R\sqrt{8 \ln 2} \approx 2.35 R$, the inverse standard deviation of the line-spread function; the signal-to-noise ratio per spectral measurement is then $\left( 1/\sqrt{2.35} \approx 0.65 \right) \times {\rm SNR}$.  
The bottom panel of Figure \ref{fig:demo}
shows the best-fit spectrum, the blue curve from the middle panel, decomposed by the individual terms in Equation \eqref{eq:mpm}.  We have smoothed the spectra to $R=1000$ for illustration purposes.  The polynomial fit (red curve) is redder than the Solar spectrum, while including the `Rayleigh term' gives a spectrum bluer than Solar.  
In practice, the `Rayleigh term' does not accurately measure the optical depth to Rayleigh scattering.  It is strongly covariant with the free polynomial, and enables Equation \eqref{eq:mpm} to fit 
broad spectral variations in cloud and surface albedos.

\section{False Positives and Significance Thresholds}

In this section, we return to our approximation to a terrestrial planet's reflection spectrum (Equation \eqref{eq:full_atm}).
We quantify the significance of a possible detection of water, oxygen and/or ozone in its atmosphere, 
using the improvement in the $\chi^2$ parameter with the addition of a molecular absorption template to Equation \eqref{eq:mpm} (as a free parameter) to measure the evidence for that species' presence.  
In order to set the minimum $\chi^2$ needed to claim a detection, we create a series of mock atmospheres using Equation \eqref{eq:full_atm}, but {\it without} a given species.  We then measure the distribution of differences in $\chi^2$ with and without constraining that species' $N_s$ to zero in Equation \eqref{eq:mpm}.  This is the distribution of $\chi^2$ improvements under the null hypothesis, and allows us to set thresholds for a given false positive probability.

\begin{figure}[t]
\vspace{-0.035\linewidth}
\noindent
\includegraphics[width=\linewidth]{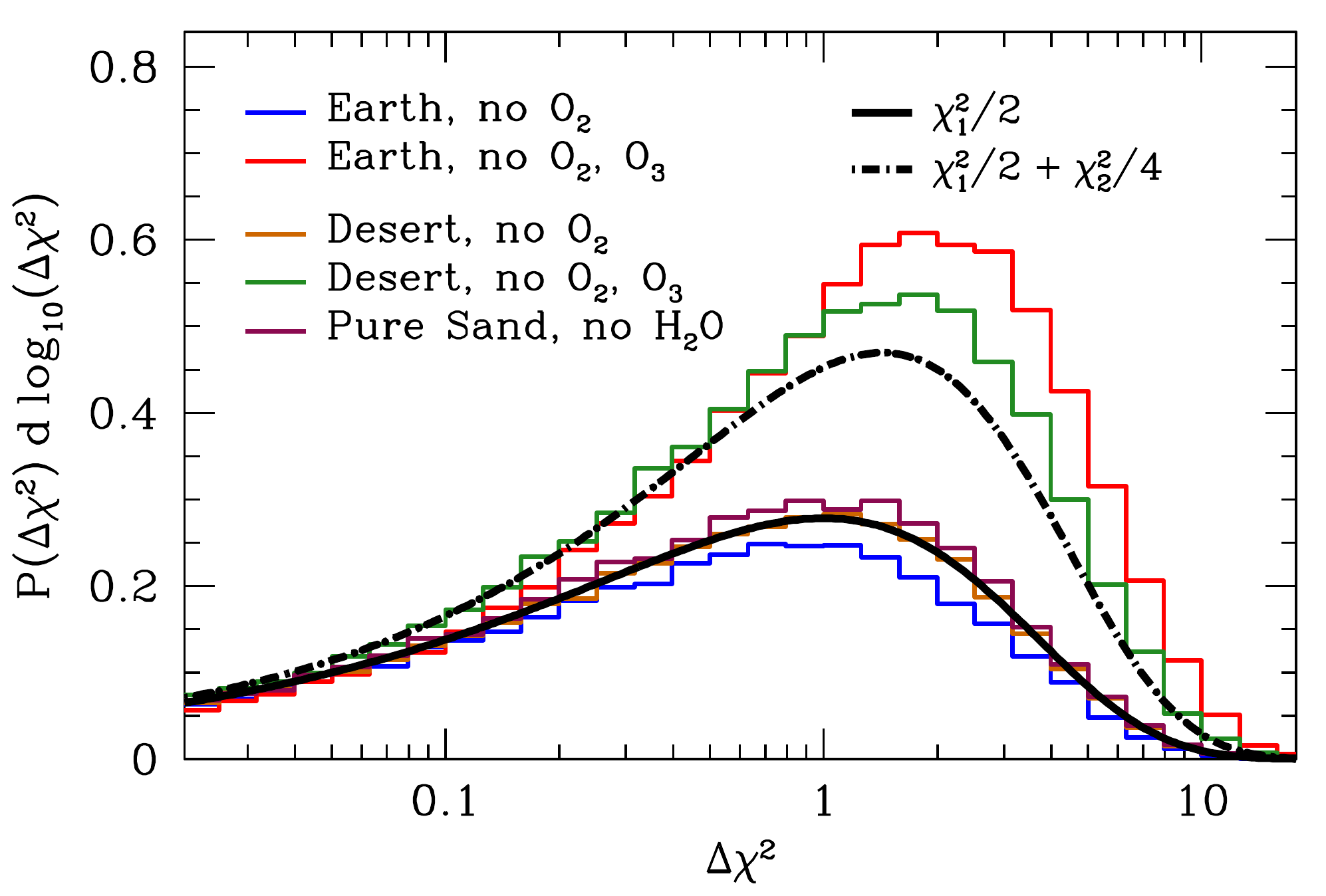}
\caption{Distributions of the improvements in $\chi^2$ by adding an additional degree of freedom represented by a species {\it not} present in a model atmosphere; the vertical axis is the probability density per logarithmic $\chi^2$, allowing the curves to be integrated by eye.  The O$_2$ and H$_2$O fits are consistent with $\chi^2_1/2$, half of the $\chi^2$ distribution with one degree of freedom (the factor of 1/2 arising because column densities must be nonnegative); the O$_2$$+$O$_3$ fits disagree somewhat with the relevant linear combination of $\chi_1^2$ and $\chi^2_2$ (the $\chi^2$ distribution with two degrees of freedom) in the tails.  We use the $\chi_1^2/2$ distribution to set $\Delta\chi^2$ detection thresholds for O$_2$ and H$_2$O. 
\label{fig:chi2_null}
}
\end{figure}

Figure \ref{fig:chi2_null}
shows the results of this test for our Earth analog with 50\% cloud cover, first setting the O$_2$ column to zero, and then setting both O$_2$ and O$_3$ to zero.  We perform the same test on our desert world assuming a 30\% cloud cover, and finally with H$_2$O on a dry exoplanet with a surface composed of 100\% sand and rock.  In the cases of O$_2$ and H$_2$O, the addition of an (unwarranted) extra degree of freedom produces a distribution of improvement in $\chi^2$ values that matches $(\chi_1^2+\delta)/2$, one-half the $\chi^2$ distribution with one degree of freedom plus a Dirac delta function (this is because of the nonnegativity constraint on the column densities: half of the distribution is a delta function at zero).  
The distribution in the case of two missing atmospheric components, O$_2$ and O$_3$, is not quite a linear combination of $\chi^2_1$ and $\chi^2_2$, the $\chi^2$ distributions with one and two degrees of freedom, particularly in the tail.  
Ozone's $\sim$0.6 $\mu$m band is a broad and shallow spectral feature that can combine with the Rayleigh term and polynomial term in Equation \eqref{eq:mpm} to reproduce a wider range of spectral shapes.

In the case of O$_2$ and H$_2$O, we use $(\chi^2_1 + \delta)/2$, $1/2$ the $\chi^2$ distribution with one degree of freedom plus a Dirac delta function, to establish our false positive thresholds.  This distribution has 99.9\% of its integrated probability below $\Delta \chi^2 = 9.6$, and 99.99\% below $\Delta \chi^2 = 13.8$; we adopt these as our thresholds for a 10$^{-3}$ and 10$^{-4}$ false positive rate, respectively.  
With two molecular species, the thresholds for a 10$^{-3}$ and 10$^{-4}$ false positive rate would become 11.8 and 16.3, respectively, if they were described by the $\chi_1^2/2+\chi^2_2/4+\delta/4$ distribution.  
The actual distributions, the red and green histograms in Figure \ref{fig:chi2_null}, have longer tails.  The $\chi^2$ values containing 99.9\% and 99.99\% of these distributions are not 11.8 and 16.3, but rather 14 and 19 for the Earth twin, and 12.4 and 17.5 for the desert world.

We also note that the addition of a second free parameter to describe H$_2$O, which significantly improves the fits in the top panel of Figure \ref{fig:demo}, changes the relevant $\chi^2$ distribution to $\chi^2_1/4 + \chi^2_2/4 + \delta/2$ (both parameters are subject to nonnegativity constraints).  The additional parameter thus modifies the $\chi^2$ thresholds to 11.4 and 16 for false positive probabilities of 10$^{-3}$ and 10$^{-4}$, respectively, illustrating the drawback of fitting a more complex model than necessary.

\section{Detecting Components of the Atmosphere and Surface}

We now turn to the probability of detecting an atmospheric constituent given our mock spectra, our eight-parameter fitting routine, and an adopted false positive threshold (which we take to be either $10^{-3}$ or $10^{-4}$).  
These detection probabilities depend on the spectral resolution and noise level, so that the probability for each species resides in a two-dimensional space.  

We consider three paths through the space of resolution and noise level.  In a best-case scenario, read noise is negligible, and the instrumental resolution may be arbitrarily high with no noise penalty.  
The variance per bin then scales as $R^{-1}$, and SNR as $R^{-1/2}$.  We consider the worst-case scenario to be a read-noise-limited instrument that simply varies the dispersion, holding everything else fixed;
the SNR in this case scales as $R^{-1}$.  Finally, we consider an intermediate case in which SNR scales as $R^{-3/4}$.  We normalize all of these paths at $R=50$. 
Table 1 summarizes our results, which we discuss in detail in the following sections.

\begin{table}
{\fignumfont Table 1. \;\;Approximate Optimal Resolutions and Minimum SNRs \hfill}
\begin{tabular}{@{\extracolsep{\fill}}lccr}
Species & $\widetilde{R}$ & ~SNR$_{\widetilde{R}}$\tablenote{90\% detection probability for $10^{-3}$ false positive rate}~ & ~SNR$_{150}$$^*$ \cr 
 \hline
 H$_2$O & 40 & 7.5 & 3.3 \cr
 O$_2$ & ~150~ & 6 & 6 \cr
 Chlorophyll~ & 20 & 120 & 40 \cr 
 \hline
\end{tabular}
\end{table}

\subsection{Oxygen and Ozone}

\begin{figure*}[t]
\vspace{-0.035\linewidth}
\noindent
\includegraphics[width=0.5\linewidth]{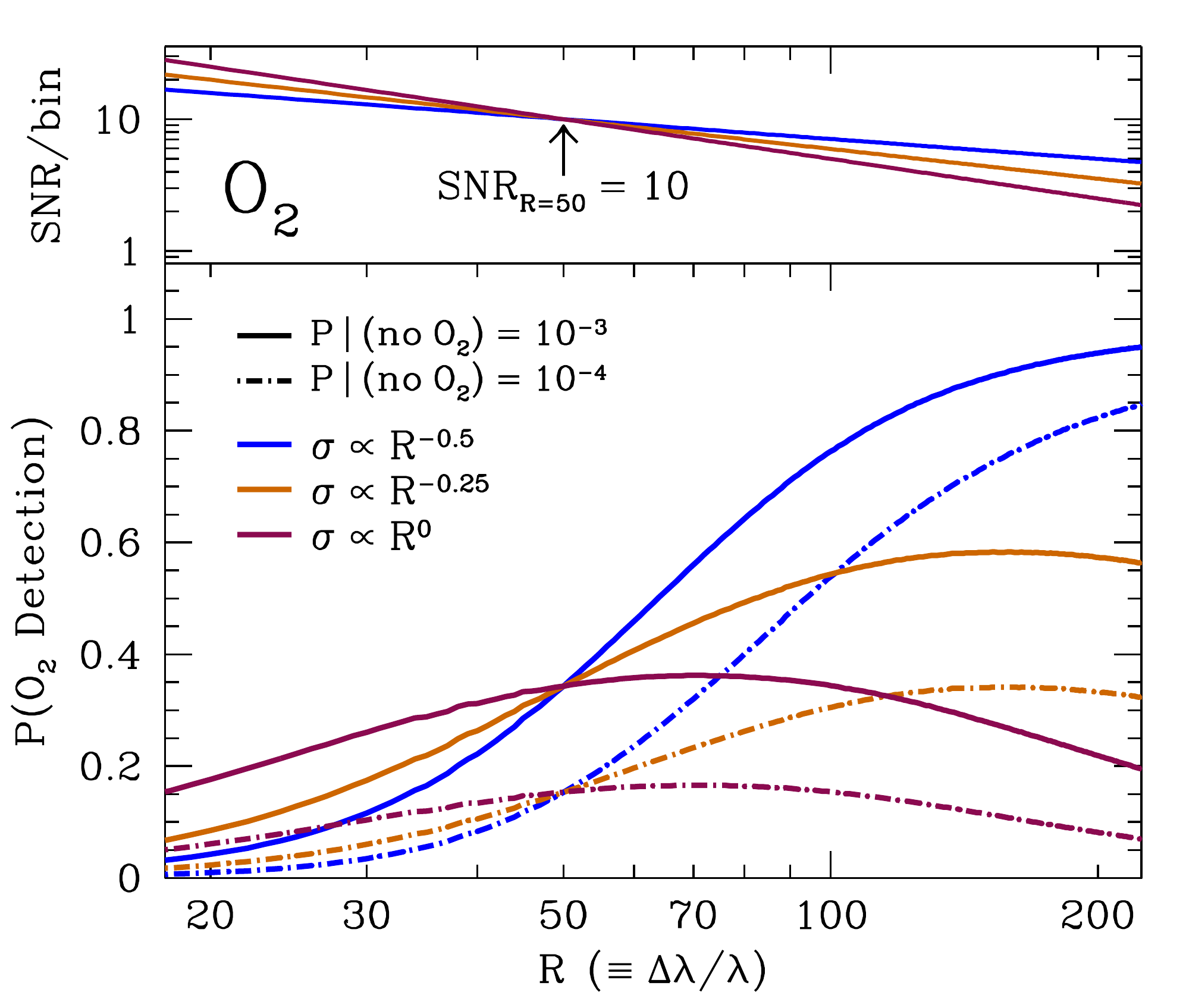}
\includegraphics[width=0.5\linewidth]{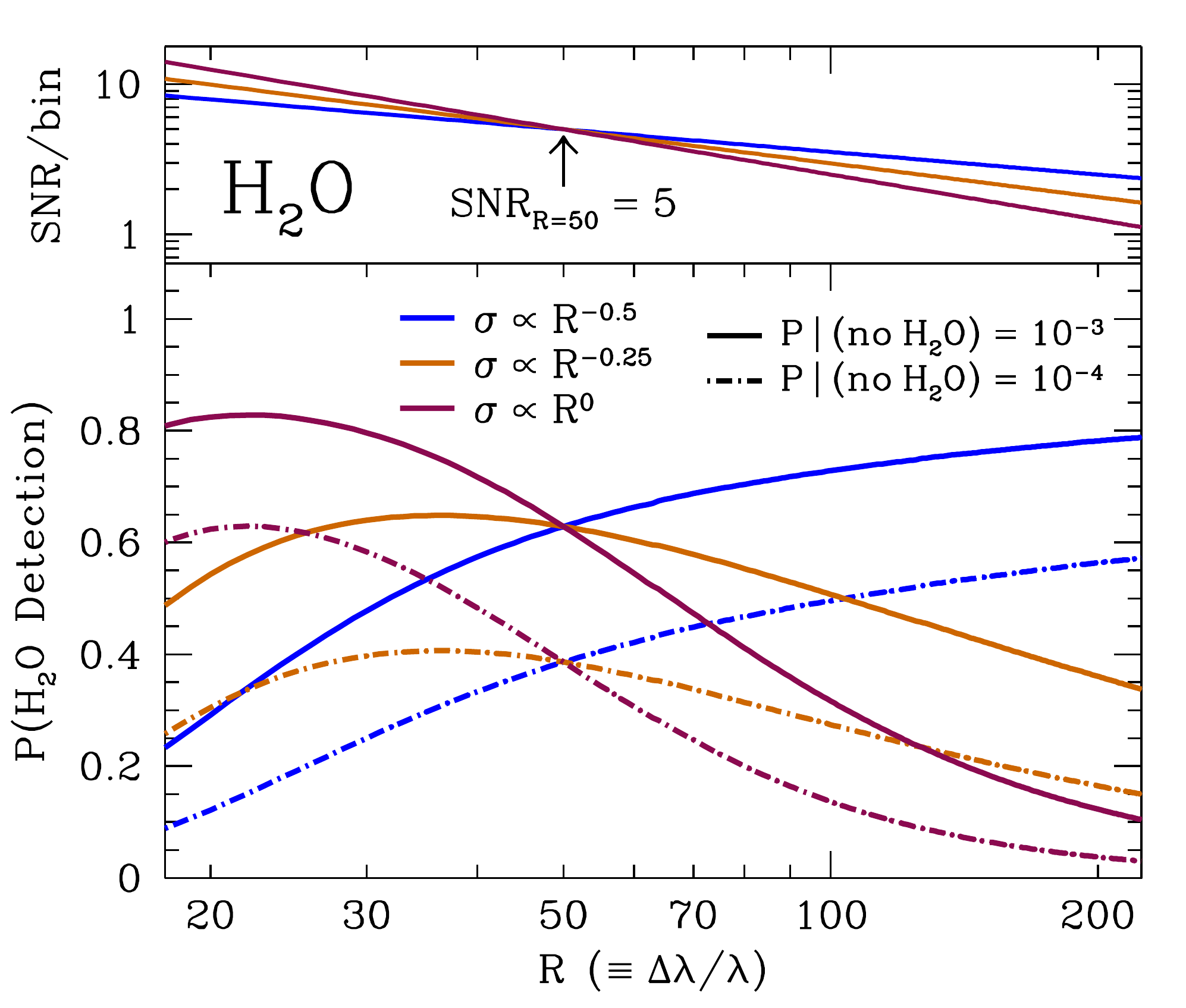}
\caption{Probability of detecting of O$_2$ (\emph{left}) and H$_2$O (\emph{right}) on an Earth twin as a function of spectral resolution and SNR.  All curves are normalized to a common SNR at $R=50$, ${\rm SNR} = 10$ for O$_2$ and ${\rm SNR} = 5$ for H$_2$O.  The curves are scaled to either reproduce the read-noise-limited case (burgundy line) in which SNR scales as $R^{-1}$, the perfect background-limited case (blue line) in which SNR scales as $R^{-1/2}$, or an intermediate case (orange line).  The solid lines indicate a false-alarm probability of $10^{-3}$, while the dot-dashed lines have a false-alarm probability of $10^{-4}$.  {\it Left:}  The optimal resolution in the intermediate case is $R\sim150$, with a corresponding minimum SNR of $\sim$5.
{\it Right:} The optimal resolution in the intermediate case is $R\sim40$, where the minimum SNR is $\sim$6.  This is a factor of $\sim$2 lower than the scaled SNR needed to detect O$_2$ in the same Earth twin.
\label{fig:P_O2H2O_Detect}
}
\end{figure*}

The left panel of Figure \ref{fig:P_O2H2O_Detect}
shows our results for the case of atmospheric O$_2$.  Given a SNR of 10 at $R=50$, the optimal resolution for an O$_2$ detection varies from $\sim$70 to many hundreds, with a value of $R\sim150$ for our intermediate noise case.  The Supporting Information contains a mostly analytic derivation of these approximate resolutions.  At $R=150$, we would need ${\rm SNR} \gtrsim 6$ for a likely detection; this corresponds to ${\rm SNR} \approx 10$ at $R=50$ in our best-case noise scaling.

Ultraviolet photons from the star will convert diatomic oxygen into ozone, so we may also ask if it would be easier to detect atmospheric oxygen in our model by simultaneously searching for both O$_2$ and O$_3$.  In the case of an Earth twin, the answer is no: the extra degree of freedom increases the $\Delta \chi^2$ threshold, and the depth of the ozone feature is insufficient to compensate (see Figure \ref{fig:demo}).  The ozone feature is also broad ($\Delta \lambda/\lambda \sim 0.1$) and occurs just redward of a major upturn in albedo from Rayleigh scattering, making it somewhat degenerate with spectrophotometric uncertainties.  This is reflected in significance thresholds somewhat higher than for the relevant $\chi^2$ distributions (as shown in Figure \ref{fig:chi2_null}).

While fitting for both O$_2$ and O$_3$ does not improve the detection probabilities for an Earth twin, it could help for a terrestrial planet with a significantly higher ozone column.  This could arise either from a higher UV flux (from an F-star, for example), from a significantly lower concentration of molecules and ions to catalyze ozone's decomposition, or both.  However, as noted earlier, any analysis simultaneously searching for O$_2$ and O$_3$ would need to set a higher significance threshold to account for ozone's ability to mimic a variable spectral albedo.

\subsection{Water}

Water has a series of deep absorption features from the red end of the visible into the near-infrared, with a variety of effective widths, making it easier to detect than diatomic oxygen.  As Figure \ref{fig:mock_spectra}
suggests, water absorption remains conspicuous in the spectrum down to spectral resolutions of $R\sim20$.  At still lower resolutions, water absorption becomes more difficult to separate from variations in the surface albedo or errors in the spectrophotometric calibration.  

The right panel of Fig.~\ref{fig:P_O2H2O_Detect}
shows the probability of a high-significance H$_2$O detection for an Earth twin, with all of the same assumptions used in the O$_2$ panel (left panel of same figure), but half of the fiducial SNR.  For the case intermediate between the optimal and pessimal noise scalings, the optimal spectral resolution for H$_2$O detection is $R\sim40$.  This is a factor of several lower than for O$_2$ and reflects the broader widths of the features.

\subsection{The Red Edge of Chlorophyll}

Chlorophyll on Earth has a sharp rise in reflectivity around 0.7 $\mu$m, the ``red edge.''  An analogous feature could be detectable on an exo-Earth, with the (large) caveat that photosynthetic extraterrestrial life may use a different family of pigment molecules than their terrestrial analogs, and the understanding that any claimed detection would be extremely controversial.  

We approximate the albedo of vegetation as a softened Heaviside step function (Equation \eqref{eq:heaviside}), which provides a reasonable match in the wavelength range from $\sim$0.5 to 1 $\mu$m.  Though the jump is very strong in pure vegetation, with the albedo increasing from $\sim$5\% to $\sim$50\%, it is much weaker in an integrated Earth spectrum.  This is due both to the small fraction of surface area covered by vegetation ($\sim$10\%), and to the fact that much of this area is covered by optically thick clouds.  
We optimistically use the same $\Delta \chi^2$ thresholds as for O$_2$ and H$_2$O to indicate a detection.  

The ``red edge'' of vegetation does not require a high spectral resolution to identify; assuming our intermediate noise scaling, a value $R\sim20$ is optimal.  Chlorophyll is, however, exceedingly difficult to detect with significance in an Earth twin.  To facilitate a comparison with O$_2$, we
explore chlorophyll's detectability as a function of SNR, the vegetation covering fraction, and the cloud fraction, at a fiducial $R$ of 150 (implicitly assuming a mission optimized to detect O$_2$).

\begin{figure}[t]
\vspace{-0.035\linewidth}
\noindent
\includegraphics[width=\linewidth]{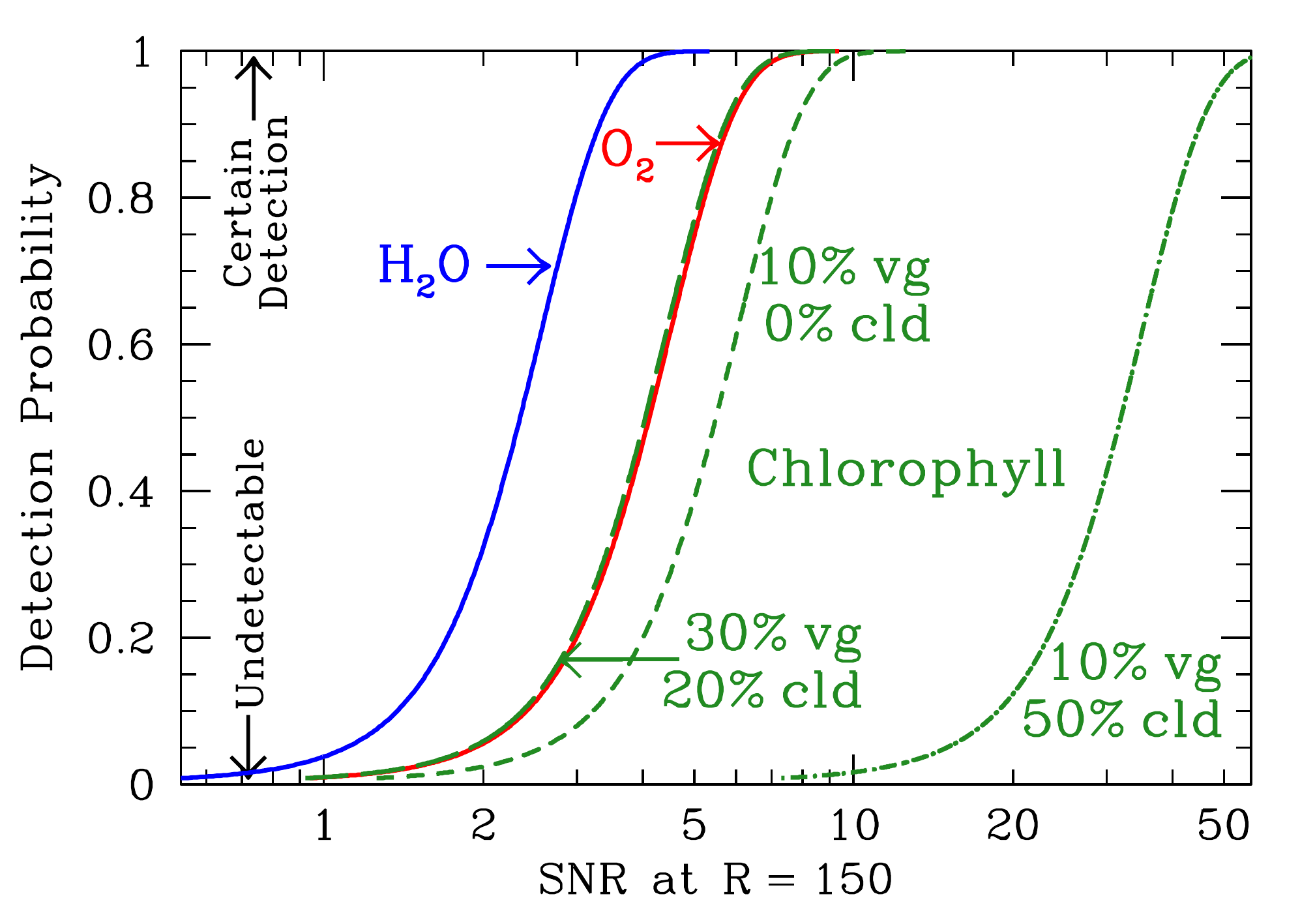}
\caption{Detectability of H$_2$O, O$_2$, and the red edge of chlorophyll for three terrestrial planets, assuming $R=150$ (i.e.~a mission optimized to detect O$_2$) and a false alarm probability of $10^{-3}$.  For an Earth twin, with 50\% cloud coverage and 10\% vegetation coverage, chlorophyll requires $\sim$6 times as much SNR as O$_2$.  Only under optimistic assumptions about cloud and vegetation cover (and the universality of chlorophyll) does the molecular family approach O$_2$ in detectability.
\label{fig:rededge}
}
\end{figure}

Figure \ref{fig:rededge} shows our results.  For an Earth twin, O$_2$ requires twice the SNR needed for H$_2$O, while chlorophyll, even if the pigment is known, requires a SNR $\sim$6 times higher than O$_2$.  At these levels, our assumption that the spectrum can be modeled with a total disregard for the details of the surface albedo begins to break down.  
In order for chlorophyll to become as easy to detect as oxygen, we must either assume a vegetation covering fraction of at least 30\% with a light cloud cover, or a cloud-free Earth.  The former scenario would have roughly half of Earth's cloud coverage and would see all land covered in lush greenery.  
The cloud-free Earth twin has a lower mean albedo, making it harder
to achieve a given SNR.  It is also difficult to imagine chlorophyll, part of a photosynthetic cycle based on water, occurring on a cloud-free world.  

While a future mission will undoubtedly search for chlorophyll on nearby terrestrial planets, we argue that a high-contrast mission should be designed to achieve the easier and better-defined goals of oxygen and water detection.  A plausible observing strategy would attempt to achieve the requisite SNR for O$_2$ and H$_2$O around nearby stars, and then spend an enormous amount of time attempting to reach the SNR needed to detect chlorophyll around the very best target(s).

\section{Conclusions}

In this paper we have constructed a minimally parametric model to recover the components of a terrestrial planet's atmosphere as observed by a future high-contrast space mission.  We find that we can reproduce the spectrum of an Earth twin to a very high accuracy even when completely neglecting the surface albedo, apart from an overall multiplicative term quadratic in wavelength.  Such a term also includes uncertainties in the spectrophotometric calibration, which are likely to be significant.  

We have focused our analysis on the optical and very near-infrared spectrum.  A Solar-type star is brightest at these wavelengths, giving the maximum photon flux.  Diatomic oxygen and water have very prominent absorption features from $\sim$0.6 to 1 $\mu$m, while likely surface materials like rock, sand, and water have nearly featureless spectral albedos.  By targeting shorter wavelengths, we also have the advantage of a finer diffraction-limited resolution.

We find that a future space mission will be likely to detect water on an Earth twin with a spectral resolution of $R \gtrsim 40$ and a SNR per bin of $\gtrsim$7.  
Such a mission will have a much more difficult time detecting atmospheric oxygen, and is unlikely to improve its sensitivity by searching for O$_2$ and O$_3$ simultaneously, at least at visible wavelengths (ozone has a strong absorption edge in the near-ultraviolet, at $\sim$0.3 $\mu$m).  
For a mission targeting only O$_2$, we find an optimal resolution of $R \sim 150$ for our intermediate noise scaling case, and a minimum SNR of $\sim$6 at $R = 150$.
This is $\sim$3 times the resolution of an instrument optimized to see water, and a factor of $\sim$2 more challenging than water as measured by the scaled SNR.

Finally, we show that the ``red edge'' of chlorophyll absorption at $\lambda \sim 0.7$ $\mu$m will be extremely difficult to detect, unless the cloud cover is much lower and/or the vegetation fraction is much higher than on Earth.  Assuming extraterrestrial chlorophyll to have the same optical properties as the terrestrial pigments, and assuming Earth-like cloud and vegetation coverings,
detecting chlorophyll will require a SNR $\sim$6 times higher than for diatomic oxygen, equivalent to a ${\rm SNR} \gtrsim 100$ at $R\sim20$.  The detectability only approaches that of O$_2$ if the cloud covering is zero, or if cloud cover is light and a much larger surface fraction, $\sim$30\%, is covered in vegetation.  

Based on our findings, we argue that a future mission should be designed towards the well-defined goal of sensitivity to O$_2$ and H$_2$O around the best candidate terrestrial exoplanets, perhaps even with two dispersing elements to achieve both $R \sim 40$ and $R \sim 150$.  Extensive (and expensive) follow-up of the very best targets, preferably with O$_2$ and H$_2$O detections, might then be used to search for the red edge of chlorophyll.

\begin{acknowledgments}
The authors acknowledge very helpful discussions with Michael McElwain and Edwin Turner, and helpful suggestions from two anonymous referees.
TDB gratefully acknowledges support from the Corning Glass Works Foundation through a membership at the Institute for Advanced Study.
DSS gratefully acknowledges support from the Association of Members of the Institute for Advanced Study.
\end{acknowledgments}

{\small 

\bibliography{refs}

}

\vspace{17 truein} 


\section{\Large Supporting Information}

\vspace{1 truein}

\section{Albedos and the Observed Flux Density}

In the main text, we deal exclusively with the reflected flux density received from an Earth-like exoplanet, which is 
the product of the incident stellar flux density and the planet's spectral albedo.  
The spectral albedo carries all of the information on composition, chemistry, and possible exobiology.  In this section, we show how the commonly plotted spectral albedo relates to the reflected flux density, and also how the albedo itself depends on the surface composition over a broader range of wavelengths than that considered in the main text.  

\begin{figure}
\centering
\includegraphics[width=\linewidth]{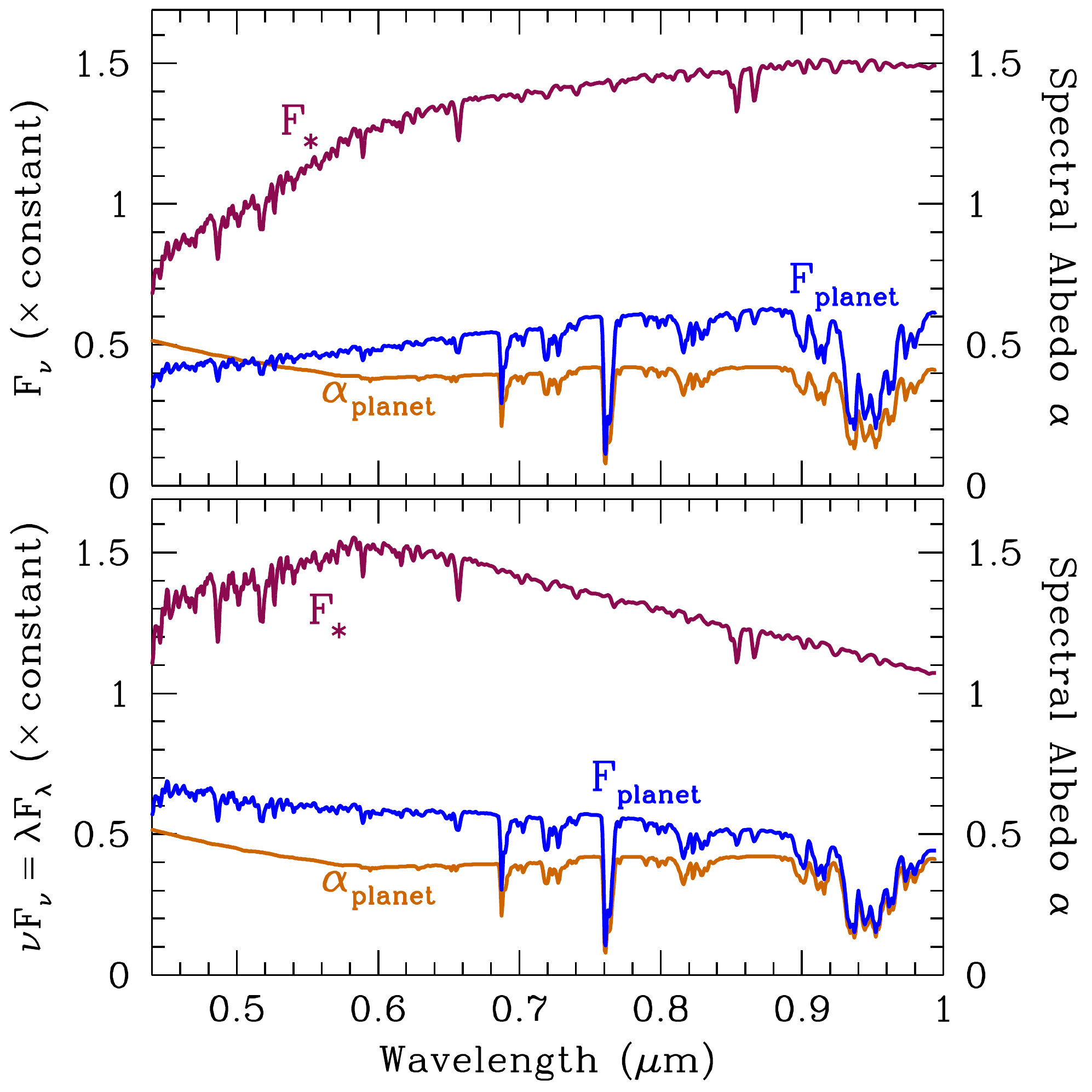}
\caption{Top panel: spectral albedo $\alpha$ (orange curve) and flux density $F_\nu$ (blue curve) for our fiducial Earth-twin mock spectrum (Equation (2) in the main text).  The stellar spectrum (burgundy curve) is the Solar spectrum at the top of the atmosphere.  Rayleigh scattering accounts for the strong upturn in albedo blueward of $\sim$0.5 $\mu$m, which compensates a downturn in the stellar flux density.  Bottom panel: the same curves, in units of $\nu F_\nu$ ($= \lambda F_\lambda$).  While these units may be more natural, we use $F_\nu$ in the main text because it is proportional to the photon flux per logarithmic unit wavelength, which corresponds to the count rate per pixel on a plausible constant-dispersion spectrograph.
\label{fig:spec_albedos}
}
\end{figure}

The top panel of Figure \ref{fig:spec_albedos} shows the (scaled) reflected flux density $F_\nu$ (blue curve) of our Earth-twin mock spectrum (Equation (2) from the main text), together with the same model's spectral albedo $\alpha$ (orange curve) and its incident stellar flux density (burgundy curve).  The Solar spectrum\footnote{http://rredc.nrel.gov/solar/spectra/am0/} falls off sharply towards the blue, with the flux density $F_\nu$ at 0.44~$\mu$m being just $\sim$55\% of its value at 0.6~$\mu$m.  The spectral albedo has a strong contribution from Rayleigh scattering at short wavelengths, raising $\alpha$ by $\sim$35\% from 0.6~$\mu$m to 0.44~$\mu$m.  These two color dependencies nearly cancel out, resulting in a planet spectrum significantly grayer than the stellar spectrum over this wavelength range.

The bottom panel of Figure \ref{fig:spec_albedos} shows the same spectra as the top panel, but in units of $\nu F_\nu$ ($= \lambda F_\lambda$).  While these units are arguably more natural, we have used $F_\nu$ in the main text because it is proportional to the photon flux $F_\gamma$ per logarithmic wavelength interval:  
\begin{align}
\frac{dF_\gamma}{d\ln \lambda} 
&= -\frac{dF_\gamma}{d\ln \nu} \nonumber \\
&= -\frac{1}{h\nu} \frac{dF}{d\nu} \frac{d\nu}{d\ln \nu} \nonumber \\
&= -\frac{1}{h} \frac{dF}{d\nu} \nonumber \\
&= -\frac{1}{h} F_\nu \propto
F_\nu~.
\end{align}
Photon flux per logarithmic wavelength interval is proportional to the count rate per pixel on a plausible spectrograph with constant dispersion.

\begin{figure}
\centering
\includegraphics[width=\linewidth]{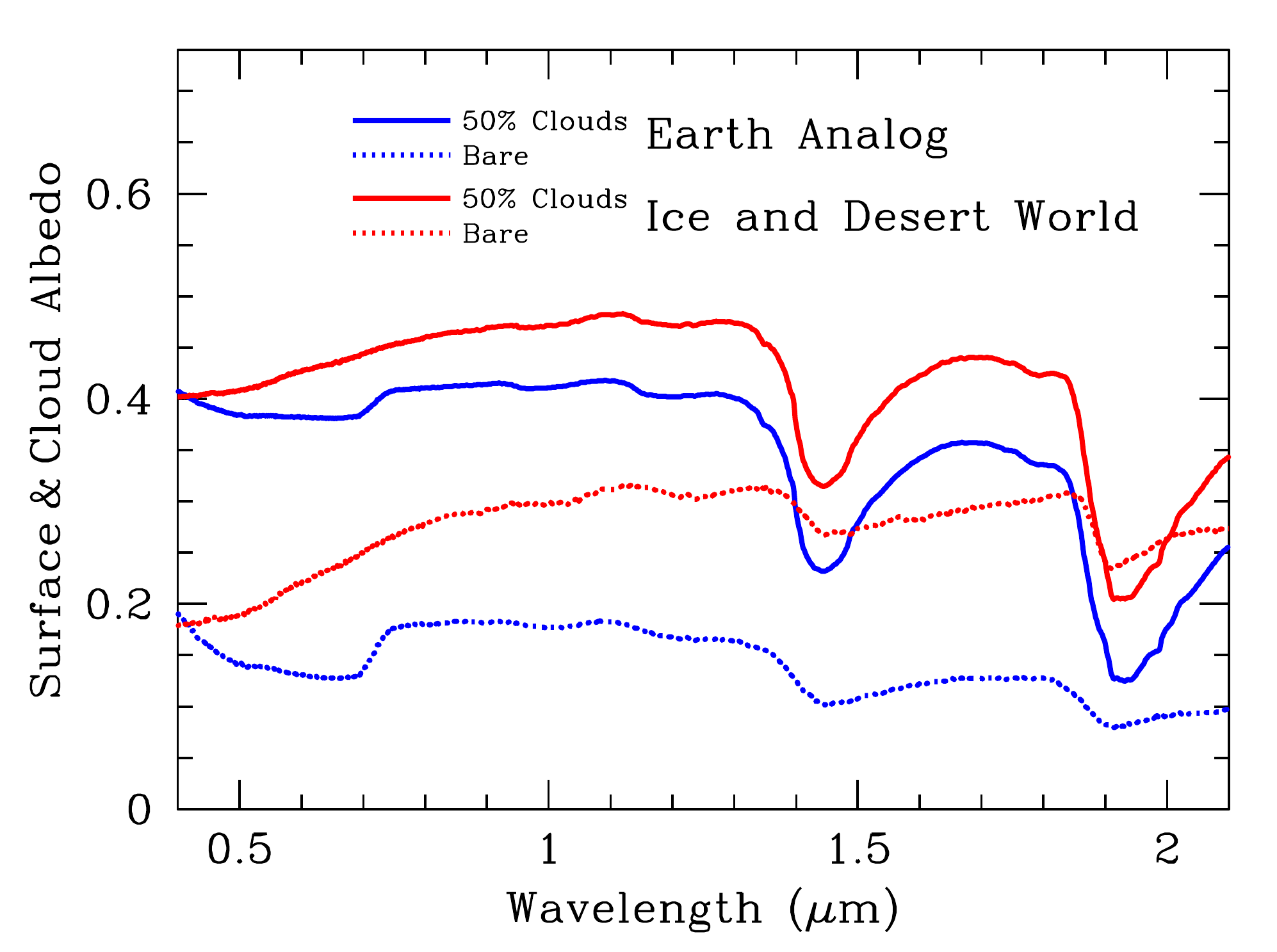}
\caption{Composite surface albedos for an Earth twin (blue curves: 10\% vegetation, 10\% sand, 70\% water, 5\% snow, 5\% dry grass) and a desert world (red curves: 60\% sand, 30\% water, and 10\% snow), each with no cloud cover (dotted lines) and a 50\% cloud cover (solid lines).  The only noticeable feature from $\sim$0.5 to 1 $\mu$m is the $0.7$ $\mu$m ``red edge'' of chlorophyll, while rocks and clouds both have wide, deep spectral features redward of 1 $\mu$m. 
\label{fig:surface_albedos}
}
\end{figure}

The effective spectral albedo of a planet, the ratio of reflected to incident flux density, is a function of the planet's surface, atmosphere, and clouds.  Figure \ref{fig:surface_albedos} shows the effective surface \& cloud albedos, neglecting atmospheric contributions, of two planets: an Earth-twin (blue curves), and a desert world (red curves).  As in the main text, our Earth twin has a surface composed of 70\% water, 10\% vegetation, 10\% sand and bare soil, 5\% snow, and 5\% dry grass.  The desert world's surface is 60\% sand, 30\% water, and 10\% snow.  Both planets are shown with zero (dotted curves) and with 50\% (solid curves) cloud covers.

The spectral albedos of sand, clouds, and water are all relatively featureless from $\sim$0.4~$\mu$m to $\sim$1.3~$\mu$m.  Both clouds and rock, however, have strong spectral features at $\sim$1.4~$\mu$m and $\sim$1.9~$\mu$m.  These would make it very difficult to model the spectrum using atmospheric constituents and low-order polynomials, as we do in the main text.  The only conspicuous feature in the visible is the red edge of chlorophyll at $\sim$0.7~$\mu$m.  While we focus on the cloudy Earth twin in the main text, our results for the detectability of atmospheric components would be almost identical for the desert world.

\section{The Optimal Instrumental Resolution for an Isolated Spectral Feature}

In the case of a top-hat spectral feature and top-hat instrumental line-spread function, the optimal spectral resolution is equal to the feature's width.  The optimal sampling in such a scenario is equivalent to binning.  Finer spectral resolution and/or finer sampling adds read noise, while coarser sampling adds background noise.  

These results do not generalize to other line/feature and filter shapes.  In this supplemental information section, we consider the more realistic case (used in the text) of a Gaussian instrumental profile.  Using several assumptions about the relative importance of read noise and background photon noise, we derive the optimal instrumental resolution in the cases of a single Gaussian and a double Gaussian feature.  We could treat other profiles (e.g.~Voigt, Lorentzian), in the same way, albeit with much more complicated algebra.  In order to keep the calculations as analytic as possible, we assume perfect knowledge of the continuum, negligible photon noise from the source itself, and a complete lack of nearby features.  The assumption of a known continuum makes the estimation of absorption and emission features equivalent.  The double Gaussian case is a particularly good approximation to the O$_2$ features, and provides a mostly analytic argument for the approximate spectral resolutions derived in the main text.

\subsection{A Single Gaussian Feature}

As in the main text, we use the instrumental resolution $R_I$ to refer to the dimensionless full width at half maximum (FWHM) of the Gaussian line-spread function (LSF),
\begin{equation}
{\rm LSF}[\lambda/\lambda_0] = \frac{R_I\sqrt{8 \ln 2}}{\sqrt{2\pi}} \exp \left[ -(4\ln 2) R_I^2 
\left( \frac{\lambda}{\lambda_0} - 1 \right)^2 \right]~,
\label{eq:linespreadprofile}
\end{equation}
for $R_I \gg 1$.  
We first consider a single Gaussian feature, with a functional form identical to Equation \eqref{eq:linespreadprofile} but in absorption, and multiplied by an arbitrary normalization $A$:
\begin{equation}
\xi [\lambda/\lambda_0] = 1 - A \frac{R_0\sqrt{8 \ln 2}}{\sqrt{2\pi}} \exp \left[ -(4\ln 2) R_0^2 
\left( \frac{\lambda}{\lambda_0} - 1 \right)^2 \right]~,
\label{eq:absline}
\end{equation}
where the continuum level is set to unity.  Throughout this analysis we will neglect uncertainties on the continuum.  Formally, we are then fitting $1 - \xi$, the continuum minus the absorption spectrum, equivalent to fitting a Gaussian feature in emission.

We consider the equivalent emission feature of unit area expected on the detector, which we denote by ${\Phi}_\lambda$.  This is the convolution of the instrumental line-spread profile (FWHM $\lambda_0/R_I$) and the normalized intrinsic feature profile (FWHM $\lambda_0/R_0$),
\begin{align}
\Phi_\lambda &= \left( \frac{1 - \xi}{A} \right) \otimes {\rm LSF} \nonumber \\
&= \frac{R_f\sqrt{8 \ln 2}}{\sqrt{2\pi}} \exp \left[ -(4\ln 2) R_f^2 
\left( \frac{\lambda}{\lambda_0} - 1 \right)^2 \right]~,
\label{eq:gauss_convolve}
\end{align}
where the effective resolution $R_f$ is 
\begin{equation}
R_f = \left( R_I^{-2} + R_0^{-2} \right)^{-1/2}~.
\end{equation}
Equation \eqref{eq:gauss_convolve} results from the fact that the convolution of two Gaussians is itself a Gaussian.

If the errors at neighboring wavelength measurements are independent, as is always the case for photon noise, the optimal estimator for the feature strength uses ${\Phi}_\lambda$, the (known) spectral profile indicated by Equation \eqref{eq:gauss_convolve}.
As an aside, we note that errors at neighboring spectral measurements need not be independent in a real spectrograph.  Complicated image-processing may be needed to derive a one-dimensional background-subtracted spectrum from a two-dimensional image, and this could easily give a non-diagonal covariance matrix.  

The observed continuum-subtracted flux density $F_\lambda$ is proportional to ${\Phi}_\lambda$ plus noise.  Suppose the feature's wavelength and width are known, and we wish to estimate $A$, the depth of the feature (from Equation \eqref{eq:absline}.  Our model is then 
\begin{equation}
F_\lambda = A {\Phi}_\lambda + {\rm noise}~.
\end{equation}
We can now use $\chi^2$ to obtain the maximum likelihood value of $A$ and its error:
\begin{equation}
\chi^2 = -2 \ln {\cal L} = \sum_\lambda \frac{\left( F_\lambda - A {\Phi}_\lambda \right)^2}{\sigma^2_\lambda}~,
\label{eq:maxlike}
\end{equation}
where ${\cal L}$ is the likelihood, the sum is over the sampled wavelengths, and $\sigma^2_\lambda$ represents the variance (assumed to be Gaussian) at each sampled wavelength.  Equation \eqref{eq:maxlike} neglects the fact that $A$ is bounded from both above and below due to the constraints that $A \geq 0$ and $\xi \geq 0$; this makes little difference unless the signal-to-noise ratio of the feature is low.
The minimum of Equation \eqref{eq:maxlike} (equivalently the maximum likelihood) occurs at
\begin{equation}
\langle A \rangle = 
\left( \sum_\lambda \frac{F_\lambda {\Phi}_\lambda}{\sigma^2_\lambda} \right)
\left( \sum_\lambda \frac{{\Phi}_\lambda^2}{\sigma^2_\lambda} \right)^{-1}~.
\end{equation}
Finally, we compute the variance of our estimate on $A$,
\begin{align}
\sigma^2_{A} &= \sum_\lambda \left(\frac{\partial \langle A \rangle}{\partial F_\lambda}\right)^2 \sigma^2_\lambda \nonumber \\
&= \left( \sum_\lambda \frac{{\Phi}_\lambda^2}{\sigma^2_\lambda} \right)^{-1}~.
\label{eq:var_A_gen}
\end{align}
We wish to maximize the signal-to-noise ratio of our measurement of $A$ or, equivalently, to minimize its noise (square root of Equation \eqref{eq:var_A_gen}).  For convenience, we use the inverse of Equation \eqref{eq:var_A_gen},
\begin{equation}
\sigma_A^{-2} = \sum_\lambda \frac{\Phi_\lambda^2}{\sigma^2_\lambda}~.
\label{eq:ivar_A_gen}
\end{equation}
This is the inverse variance of our measurement of $A$, which is proportional to the square of the signal-to-noise ratio.  
In the next section, we will seek to maximize it by varying $R_I$, the instrumental resolution.

\subsection{The Optimal Resolving Power}

The optimal resolving power minimizes the variance on our measurement of the intrinsic feature strength (given by Equation \eqref{eq:var_A_gen}).  
To make the problem analytic, we assume that the spectra are well-sampled (practically, this means Nyquist or better).  This allows us to write the sums as integrals.  The read noise per unit wavelength scales as $R_I$; the photon noise per unit wavelength is independent of $R_I$.  We also introduce the concept of $R_{\rm eq}$, the instrumental resolution at which read and photon noise contribute equally. 
The ratio $\sigma^2_r/\sigma^2_b$ is thus proportional to $R_I$ and is unity when $R_I = R_{\rm eq}$.  The total variance per unit wavelength is then
\begin{align}
\sigma^2_\lambda \Delta \lambda &= \sigma^2_r + \sigma^2_b \nonumber \\
&= \sigma^2_b \left( \frac{\sigma^2_r}{\sigma^2_b} + 1 \right) \nonumber \\
&= \sigma^2_b \left( \frac{R_I}{R_{\rm eq}} + 1 \right) \nonumber \\
&\propto \frac{R_I}{R_{\rm eq}} + 1~.
\label{eq:noise_balance}
\end{align}
Equation \eqref{eq:noise_balance} neglects any wavelength dependence of the background noise, a reasonable assumption for a narrow spectral feature.
We then invert Equation \eqref{eq:var_A_gen} to obtain Equation \eqref{eq:ivar_A_gen} (avoiding the negative exponent) and, replacing sums with integrals, we have
\begin{align}
\sigma_A^{-2} &\propto \int_0^\infty \frac{R_f^2 d\lambda}{R_I/R_{\rm eq} + 1}
\exp \left[ -(8 \ln 2) R_f^2 \left(\frac{\lambda}{\lambda_0} - 1 \right)^2 \right] \nonumber \\
&\approx \frac{R_f}{R_I/R_{\rm eq} + 1} \int_{-\infty}^\infty 
\frac{dx}{\sqrt{8 \ln 2}} \exp\left[-x^2 \right] \nonumber \\
&\propto \left( R_I^{-2} + R_0^{-2} \right)^{-1/2} \left( R_I/R_{\rm eq} + 1 \right)^{-1}~,
\label{eq:inv_var_A}
\end{align}
where 
\begin{equation}
x = R_f\sqrt{8 \ln 2} \left(\frac{\lambda}{\lambda_0} - 1 \right)
\end{equation}
and we have used the fact that $R_f \gg 1$ to set the limits of integration to $\pm \infty$.  Maximizing Equation \eqref{eq:inv_var_A} is equivalent to minimizing Equation \eqref{eq:var_A_gen}, and gives the remarkably simple solution 
\begin{equation}
R_I = \left( R_0^2 R_{\rm eq} \right)^{1/3}~,
\end{equation}
where $R_0$ is the intrinsic FWHM of the feature and $R_{\rm eq}$ is the resolution of read/background noise equality.  The resolution $R_I = R_0$ is optimal only in the case where $R_0$ is also the resolution where read noise and background noise are equal.  If read noise dominates, a lower resolution is preferred, while if photon noise dominates, a higher resolution is optimal.  In the limit of zero read noise, $R_{\rm eq} \rightarrow \infty$, and the optimal resolution of the instrument increases without bound.  

\subsection{A Sum of Two Gaussians}

We now generalize the above analysis to the case where the spectral feature is the sum of two overlapping Gaussians, with the relative normalization of the Gaussians known and fixed.  As we will show, this actually provides a very good approximation to the O$_2$ features at 0.76 and 0.69 $\mu$m.  We assume that the two Gaussians have normalizations $A_1$ and $A_2$ (such that $A_1/A_2$ is the ratio of areas) and widths $R_1$ and $R_2$, and are separated by $\delta \lambda/\lambda = 1/\beta$, with $R_1$, $R_2$, and $\beta \gg 1$ (the features are all narrow).  

The spectral template for the composite feature is the convolution of the line-spread function with the sum of the two Gaussian components.  The template for the first Gaussian is
\begin{equation}
{\Phi}_1 = A_1 R_{f,1} \exp \left[ -(4 \ln 2)R_{f,1}^2 \left(\frac{\lambda}{\lambda_0} - 1 - \frac{1}{2\beta} \right)^2 \right]~,
\end{equation}
with 
\begin{equation}
R_{f,1} = \left(R_1^{-2} + R_I^{-2} \right)^{-1/2}~.
\end{equation}
and the template for the composite Gaussian is ${\Phi}_1 + {\Phi}_2$.  The inverse variance on a measurement of the feature's depth is then
\begin{align}
\sigma^{-2}_A &= \int_0^\infty \frac{d\lambda}{\sigma^2_r + \sigma^2_b}
\left( {\Phi}_1[\lambda] + {\Phi}_2[\lambda] \right)^2 \nonumber \\
&= \int_0^\infty \frac{d\lambda}{\sigma^2_r + \sigma^2_b}
\left( {\Phi}_1^2[\lambda] + {\Phi}_2^2[\lambda] + 2 {\Phi}_1[\lambda] {\Phi}_2[\lambda]\right)~.
\label{eq:2gaussmatchedfilt}
\end{align}
The first two terms in Equation \eqref{eq:2gaussmatchedfilt} integrate exactly as in the case of a single Gaussian, e.g.
\begin{equation}
\int_0^\infty \frac{d\lambda {\Phi}_1^2 [\lambda]}{\sigma^2_r + \sigma^2_b} 
\propto \frac{A_1^2 R_{f,1}^2}{1 + R_I/R_{\rm eq}}~.
\end{equation}
The cross term  is more complicated,
\begin{align}
{\Phi}_1 {\Phi}_2 &= A_1A_2 R_{f,1} R_{f,2} \exp \Bigg[-(4\ln 2) 
\Bigg( R_{1,f}^2 \left(\frac{\lambda}{\lambda_0} - 1 - \frac{1}{2\beta} \right)^2 \nonumber \\
  &\qquad\qquad\qquad\qquad\qquad\qquad\qquad + R_{2,f}^2 \left(\frac{\lambda}{\lambda_0} - 1 + \frac{1}{2\beta} \right)^2
\Bigg) \Bigg]
\nonumber \\
&= A_1A_2 R_{f,1} R_{f,2} \exp \Bigg[-(4\ln 2) 
\Bigg( \frac{R_{1,f}^2 R_{2,f}^2}{\beta^2 (R_{1,f}^2 + R_{2,f}^2)} \nonumber \\
  &\qquad\qquad\qquad\qquad\qquad + 
\left( R_{1,f}^2 + R_{2.f}^2 \right) \left(\frac{\lambda}{\lambda_0} - 1 + C \right)^2 \Bigg) \Bigg]~.
\end{align}
where 
\begin{equation}
C = \frac{1}{2\beta} \left( \frac{R_{f,2}^2 - R_{f,1}^2}{R_{f,1}^2 + R_{f,2}^2} \right) \ll 1~.
\end{equation}
After integrating, and using the fact that $R_1$, $R_2$ and $\beta \gg 1$ to set the limits to $\pm \infty$ and drop $C$, we obtain 
\begin{align}
\sigma^{-2}_A \propto \frac{1}{1 + R_I/R_{\rm eq}}
&\Bigg( A_1^2 R_{1,f} + A_2^2 R_{2,f} \nonumber \\
&+ A_1 A_2 \frac{R_{1,f} R_{2,f} \sqrt{8}}{\sqrt{R_{1,f}^2 + R_{2,f}^2}}
\exp \left[ -\frac{(4 \ln 2) R_{1,f}^2 R_{2,f}^2}{\beta^2 (R_{1,f}^2 + R_{2,f}^2)} \right]
\Bigg)~.
\label{eq:2gauss_result}
\end{align}
Notice that Equation \eqref{eq:2gauss_result} reduces to 
\begin{equation}
\sigma^{-2}_A \propto \frac{R_f \left( A_1 + A_2 \right)^2}{1 + R_I/R_{\rm eq}}
\end{equation}
in the limit $\beta \rightarrow \infty$ and $R_1 = R_2$ (a single Gaussian of amplitude $A_1 + A_2$).  

\subsection{Application to the O$_2$ Features}

\begin{figure}
\centering\includegraphics[width=\linewidth]{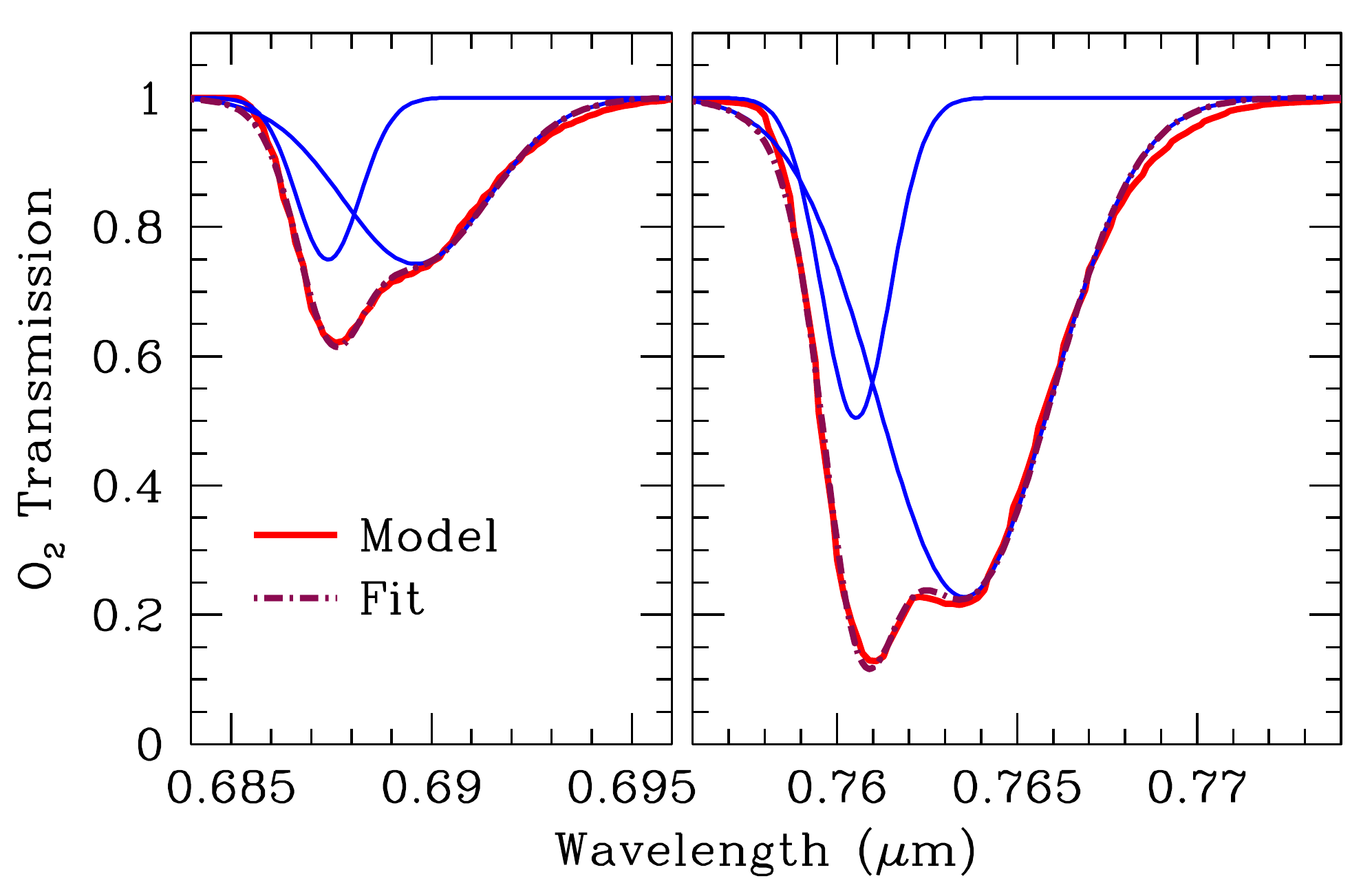}
\caption{Double Gaussian fits to the O$_2$ features at 0.69 and 0.76 $\mu$m.  We use the fitted parameters to evaluate Equation \eqref{eq:2gauss_result} as a function of the instrumental resolution.  \label{fig:o2_fit}}
\end{figure}

Unfortunately, Equation \eqref{eq:2gauss_result} cannot be solved for $R_I$ analytically.  We therefore fit the O$_2$ features at 0.76 $\mu$m and 0.69 $\mu$m in the Rugheimer et al.~(2013) absorption spectra as the sums of two Gaussians and compute Equation \eqref{eq:2gauss_result} as a function of $R_I$ for the fitted parameters of the Gaussians.  Figure \ref{fig:o2_fit} shows that a pair of Gaussians provides an excellent fit to each feature.  For the 0.69 $\mu$m band, we find $A_1/A_2 = 0.429$, $R_1 = 366$, $R_2 = 161$, and $\beta = 317$; for the 0.76 $\mu$m band we find $A_1/A_2 = 0.255$, $R_1 = 340$, $R_2 = 135$ and $\beta = 254$.

Finally, in addition to the combination of read noise and photon noise used in Equation \eqref{eq:2gauss_result}, we add the result with $\sigma^2 \propto \sqrt{R_I}$ (used in the main text as an intermediate case between the read noise and photon noise limits).  If we normalize the prefactors to have the same value at $R_{\rm eq}$, i.e.
\begin{equation}
\sigma^2_1 = (1 + R_I/R_{\rm eq}) \quad {\rm and} \quad
\sigma^2_2 = 2\sqrt{R_I/R_{\rm eq}}~,
\end{equation}
their derivatives, evaluated at $R_{\rm eq}$, are also equal.  Intuitively then, our intermediate case corresponds to the read and photon noise contributing equally at all instrumental resolutions.  

\begin{figure}
\noindent
\includegraphics[width=\linewidth]{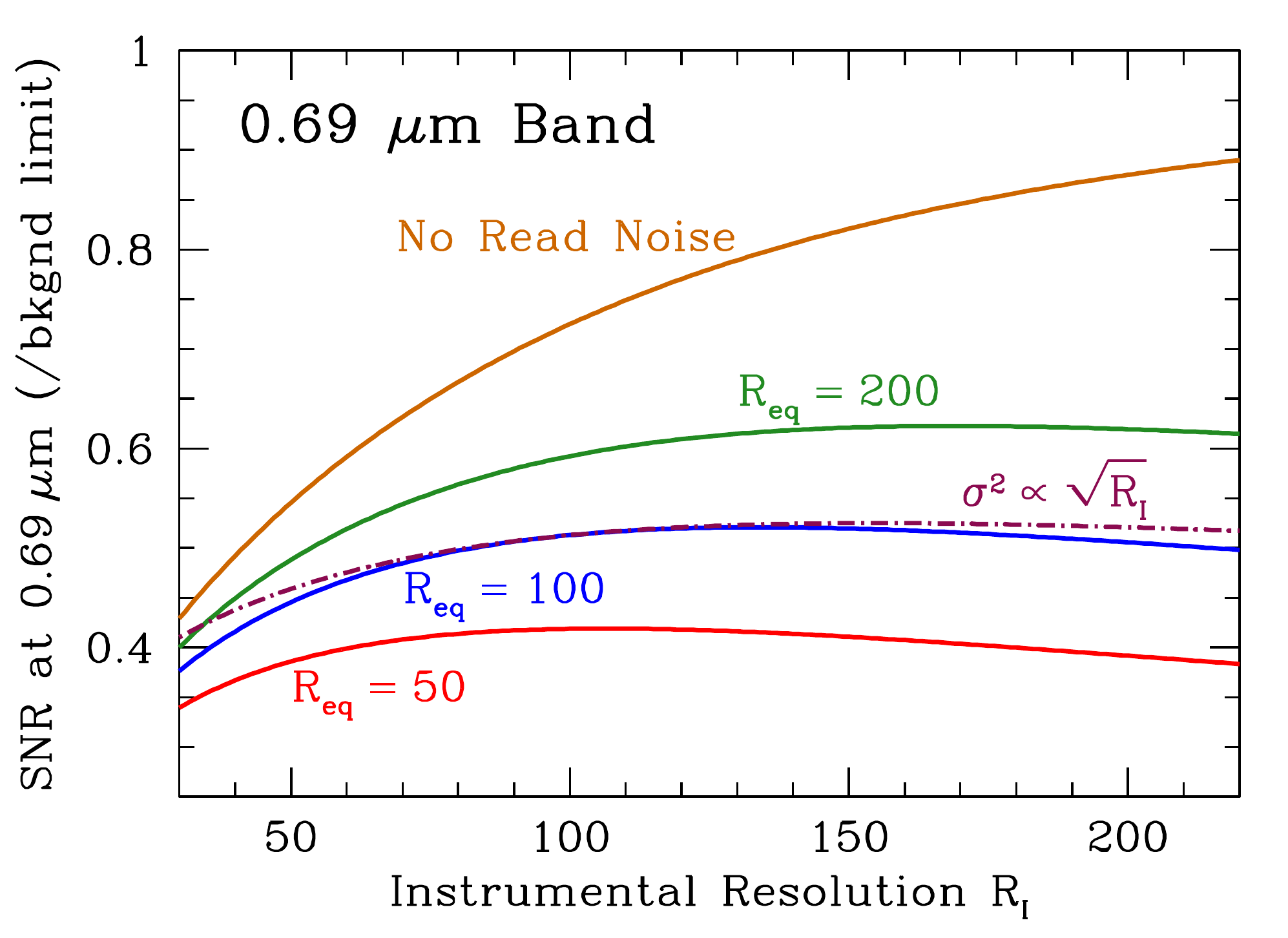}
\includegraphics[width=\linewidth]{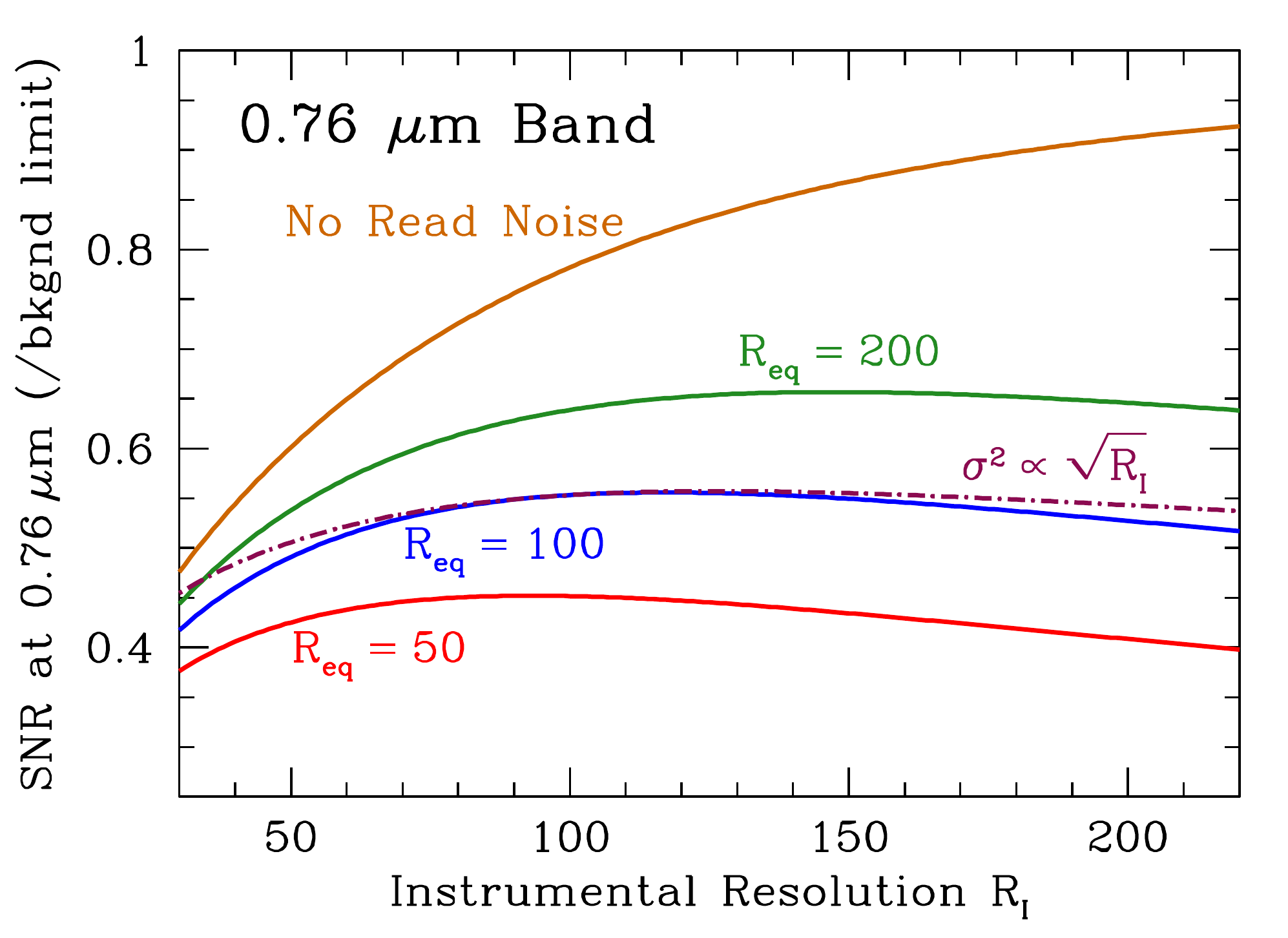}
\caption{The signal-to-noise ratio (SNR) of the 0.69 $\mu$m (left) and 0.76 $\mu$m (right) O$_2$ features relative to the infinite-resolution background-limited case.  We consider five scenarios: a read-noise-free case (orange lines), $R_{\rm eq} = 50$, 100, and 200 (the instrumental resolution at which read noise and background noise contribute equally), and the intermediate noise scaling $\sigma^2 \propto \sqrt{R_I}$ used in the main text.  The optimal $R_I$ range from $\sim$100 to $\infty$, with values $\sim$140 for our intermediate noise case.  Effects not considered here, like blending from other features and the definition of the continuum, would depress the SNR at low values of $R_I$. \label{fig:o2_snr}}
\end{figure}

Figure \ref{fig:o2_snr} shows the results of this exercise for the 0.69 $\mu$m and 0.76 $\mu$m O$_2$ bands in each of five scenarios: no read noise (orange lines), $R_{\rm eq}$ (resolutions of background/read noise equality) of 50, 100, and 200, and the intermediate scenario we use in the main text, with $\sigma^2 \propto \sqrt{R_I}$.  We normalize this intermediate scenario to the $R_{\rm eq} = 100$ curve at $R_I = 100$.  

In all cases, and even for $R_{\rm eq} = 50$, the optimal resolution is higher than the canonical $R = 70$ width of the oxygen feature.  For our intermediate noise scenario ($\sigma^2 \propto \sqrt{R_I}$), the optimal instrumental resolutions at 0.69 $\mu$m and 0.76 $\mu$m are $\sim$160 and $\sim$130, which agree nicely with the values computed in the main text.  A reviewer of this manuscript pointed out that the widths of the 0.69 $\mu$m and 0.76 $\mu$m O$_2$ features published in 
\cite{DesMarais+Harwit+Jucks+etal_2002}, 
$R=54$ and $R=69$, respectively, disagree with Figure \ref{fig:o2_fit}, and appear to be in error.  Figure \ref{fig:o2_fit}, shown at a spectral resolution $R \sim 400$, indicates full widths at half maximum of corresponding to $R \sim 160$ and $R \sim 115$ for the 0.69 $\mu$m and 0.76 $\mu$m features, respectively.

The simplified calculations in this supporting material neglect the measurement of the continuum and confusion with neighboring features.  While our analytic results do provide intuition for the main results we present in the text, they cannot be trusted in detail.  The model we present in the main text fully accounts for the confusion in the continuum, the use of all spectral features from $\sim$0.5 to 1 $\mu$m, and blending between spectral features of different species.  These effects tend to depress the signal-to-noise ratio in the curves of Figure \ref{fig:o2_snr} at low $R_I$, and account for the modest differences between the optimal $R_I$ in the main text and in Figure \ref{fig:o2_snr}.

\end{article}

\end{document}